\documentclass[journal]{IEEEtran}
\usepackage{booktabs}
\usepackage{multirow}
\usepackage{amsmath}
\usepackage{amsfonts}
\usepackage{array}
\usepackage{graphicx} % picture
\usepackage{cite}
\usepackage{color} 
\begin{document}
\title{ How to Build a Graph-Based Deep Learning Architecture in Traffic Domain: A Survey}
\author{Jiexia~Ye,~Juanjuan~Zhao*,~Kejiang~Ye, IEEE~Member, ~Chengzhong~Xu, IEEE~Fellow\\
\thanks{*Corresponding author: Juanjuan Zhao}
\thanks{
 Jiexia Ye, Juanjuan Zhao, Kejiang Ye are with Shenzhen Institutes of Advanced Technology, Chinese Academy of Sciences, China (E-mail: \{jx.ye, jj.zhao, kj.ye\}@siat.ac.cn).

Chengzhong Xu is with State Key Lab of IOTSC, Department of Computer Science, University of Macau, Macau SAR, China (E-mail: czxu@um.edu.mo).
}
}

\maketitle
\begin{abstract}
In recent years, various deep learning architectures have been proposed to solve complex challenges (e.g. spatial dependency, temporal dependency) in traffic domain, which have achieved satisfactory performance. These architectures are composed of multiple deep learning techniques in order to tackle various challenges in traffic tasks. Traditionally, convolution neural networks (CNNs) are utilized to model spatial dependency by decomposing the traffic network as grids. However, many traffic networks are graph-structured in nature. In order to utilize such spatial information fully, it's more appropriate to formulate traffic networks as graphs mathematically. Recently, various novel deep learning techniques have been developed to process graph data, called graph neural networks (GNNs). More and more works combine GNNs with other deep learning techniques to construct an architecture dealing with various challenges in a complex traffic task, where GNNs are responsible for extracting spatial correlations in traffic network. These graph-based architectures have achieved state-of-the-art performance. To provide a comprehensive and clear picture of such emerging trend, this survey carefully examines various graph-based deep learning architectures in many traffic applications. We first give guidelines to formulate a traffic problem based on graph and construct graphs from various kinds of traffic datasets. Then we decompose these graph-based architectures to discuss their shared deep learning techniques, clarifying the utilization of each technique in traffic tasks. What's more, we summarize some common traffic challenges and the corresponding graph-based deep learning solutions to each challenge. Finally, we provide benchmark datasets, open source codes and future research directions in this rapidly growing field.
\end{abstract}


\section{Introduction}
\IEEEPARstart{A}{long} with the acceleration of urbanization process, mass population is quickly gathering together towards cities. In many cities, especially cities in developing countries, the rapidly increasing number of private vehicles and growing demand of public transport services are putting great pressure on their current transportation systems. The traffic problems such as frequent traffic jams,  serious traffic accidents and long commute have seriously decreased the operation efficiency of cities and degraded the travel experience of passengers. To address these challenges, many cities are committed to develop an Intelligent Transportation System (ITS) which can provide efficient traffic management, accurate traffic resources allocation and high-quality transportation service. Such a system can reduce traffic accidents, relieve traffic congestion and ensure public traffic safety.

To construct an Intelligent Transportation System which makes cities smart, there are mainly two indispensable components, i.e. intelligent infrastructures and advanced algorithms.

On one hand, with the increasing investment in transportation infrastructures, there are more and more traffic equipments and systems, including loop detectors, probes, cameras on road networks, GPS in taxis or buses, smart cards on subways and buses, automatic fare collection system and online ride-hailing system. These infrastructures produce traffic data around-the-clock, which are heterogeneous data, including numeric data (e.g. GPS trajectories, traffic measurements), image/video data (e.g. vehicle images) and textual data (e.g. incident reports). These transportation data are enormous in volume and complicated in structure,  containing complex traffic patterns (e.g. spatiotemporal dependency, highly nonlinearity,  complex dynamics). There is an urgent need to utilize more intelligent and powerful approaches to process such traffic data.

On the other hand, in transportation domain, researchers have witnessed the algorithms evolving from statistical methods, to machine learning models and recently to deep learning approaches. In the early stage, statistic methods including ARIMA and its variants \cite{DBLP:conf/icassp/YuZ04},\cite{williams2003modeling}, VAR\cite{DBLP:journals/jits/ChandraA09}, Kalman filtering \cite{xie2007short} were prevalent, as they have solid and widely accepted mathematical foundations. However, the linear and stationarity assumptions of these methods are violated by the highly non-linearity and  dynamics in traffic data, resulting in poor performance in practice. Traditional machine learning approaches such as Support Vector Machine \cite{DBLP:journals/ijon/FuMLL16}, K-Nearest Neighbors\cite{DBLP:conf/icdm/MayHKSS08} can model non-linearity and extract more complex correlations in traffic data. However, the shallow architecture, manual feature selection and separated learning in these models are considered to be unsatisfactory in big data scenarios \cite{liu2020urban}.

The breakthrough of deep learning in many domains, including computer vision, natural language processing has attracted attention from transportation industry and research community. Deep learning techniques overcome the handcrafted feature engineering by providing an end-to-end learning from raw traffic data. The powerful capacities of deep learning techniques to approximate any complex functions in theory can model more complicated patterns in various traffic tasks. In recent years, due to the increasing computing power (e.g. GPU) and sufficient traffic data \cite{liu2020urban}, deep learning based techniques have been widely employed and achieved state-of-the-art performance in various traffic applications. The Recurrent neural networks (RNNs) and Convolutional neural networks (CNNs) based architectures used to be popular in extracting spatiotemporal dependencies. In these architectures, RNN or its variants are employed to extract the temporal correlations in traffic data \cite{DBLP:conf/ijcai/LvX0YZZ18}. CNNs are used to capture the spatial correlations in grid-based traffic network \cite{ma2017learning}. However, many traffic networks are graph-structured in nature, e.g. road network \cite{DBLP:conf/infocom/YanSZXLQ17} and subway network. The spatial features learned in CNN are not optimal for representing the graph-based traffic network. Although some previous works have analyzed traffic problems in a graph view \cite{sun2017discovering},\cite{sun2014spatial}, these traditional approaches are not powerful enough to process big data and tackle complicated correlations in traffic network. 

Recently, many researchers have extended deep learning approaches on graph data to exploit graph structure information \cite{gori2005new} and proposed a new group of neural networks called graph neural networks (GNNs)\cite{scarselli2008graph},\cite{henaff2015deep},\cite{li2018learning}, which aims to address graph-related applications. GNNs have become the state-of-the-art approaches in many domains, including  computer vision \cite{chen2019multi-label}, natural language processing \cite{guo2019attention}, biology \cite{duvenaud2015convolutional}, recommendation system \cite{ying2018graph}. Since many traffic data are graph-structured, many existing works incorporate GNNs into a deep learning architecture to capture the spatial dependency. Recent works have shown that such GNNs-based architectures can achieve better performance than CNNs-based architectures, for that most traffic networks are graph-structured naturally and GNNs can extract the spatial dependency more accurately. 
In addition, some  tasks inherently require researchers to conduct prediction based on a graph, e.g. prediction in traffic network with irregular shapes.
Many related works have been produced during the last couple of years and more are on the road. Under this circumstance, a comprehensive literature review on these graph-based deep learning architectures in transportation domain would be very timely, which is exactly our work. 

To our best knowledge, we are the first to provide a comprehensive survey on graph-based deep learning works in traffic domain. Note that some works we review actually work on similar traffic problems with similar techniques. Our work can help the upcoming researchers avoid repetitive works and focus on new solutions. What's more, the practical and clear guidance in this survey enables participators to apply these new emerging approaches in real-world traffic tasks quickly.

To sum up, the main contributions of this paper are as follows:
\begin{itemize}
\item {We systematically outline traffic problems, related research directions, challenges and techniques in traffic domain, which can help related researchers to locate or expand their researches.}
\item {We summarize a general formulation about various traffic problems and provide a specific guidance to construct graphs from several typical kinds of raw traffic datasets. Such thorough summarization is quite practical and can accelerate the applications of graph-based approaches in traffic domain.}
\item{We provide a comprehensive review over typical deep learning techniques widely used in graph-based traffic works. We elaborate their theoretical aspects, advantages, limitations and variants in specific traffic tasks, hoping to inspire the followers to develop more novel models.}
\item{We discuss some challenges shared by most graph-based traffic tasks. For each challenge, we conclude multiple deep learning-based solutions and make necessary comparison, providing useful suggestions for model selection in traffic tasks.}
\item{We collect benchmark datasets, open-source codes in related papers to facilitate baseline experiments in traffic domain. Finally, we propose some future research directions.}
\end{itemize}

The rest of the paper is organized as follows. Section \ref{sec:Related} presents some surveys in traffic domain and some reviews about graph neural networks. Section \ref{sec:Problems} briefly outlines several traffic problems and the corresponding research directions, challenges and solutions. Section \ref{sec:GraphConstruction} summarizes a general formulation about traffic problems and the graph construction from traffic datasets. Section \ref{sec:Techniques} analyzes the functionality, advantages and defects of GNNs and other deep learning techniques, as well as examining the tricks to create novel variants of these techniques in specific traffic tasks. Section \ref{sec:Challenges} discusses common challenges in traffic domain and the corresponding multiple solutions. Section \ref{sec:Open} provides hyperlinks of  datasets and  open codes in papers we investigate. Section \ref{sec:Future} presents future directions. Section \ref{sec:Conclusion} concludes the paper.

\section{Related Research Surveys}
\label{sec:Related}
There have been some surveys summarizing the development process of algorithms in traffic tasks from different perspectives. Karlaftis et al. \cite{karlaftis2011statistical} discussed differences and similarities between statistical methods and neural networks to promote the comprehension between these two communities.  Vlahogianni et al. \cite{vlahogianni2014short} reviewed ten challenges on short-term traffic forecasting, which stemmed from the changing needs of ITS applications. Xie et al. \cite{xie2020urban} conducted a comprehensive overview of approaches in urban flow forecasting.  Liu et al. \cite{liu2020urban} classified deep learning based urban big data fusion methods into three categories, i.e. DL-output-based fusion, DL-input-based fusion and DL-double-stage-based fusion. Deep learning approaches for popular topics including traffic network representation, traffic flow forecasting, traffic signal control, automatic vehicle detection are discussed in \cite{nguyen2018deep}, \cite{wang2019enhancing}. Veres et al. \cite{veres2019deep} and Chen et al.\cite{chen2019survey} gave a similar but more elaborate analysis on new emerging deep learning models in various transportation topics. Wang et al. \cite{wang2019deep} provided a spatial-temporal perspective to summarize deep learning techniques in traffic domain and other domains. However, all these surveys do not take graph neural networks (GNNs) related literatures into consideration, except that Wang et al. \cite{wang2019deep} mentioned GNNs but in a very short subsection. 

On the other hand, in recent years, there are several reviews summarizing literatures about GNNs in different aspects. Bronstein et al. \cite{bronstein2017geometric} is the first to overview deep learning techniques on processing data in non-Euclidean space (e.g. graph data). Zhou et al. \cite{zhou2018graph} categorized GNNs into graph types, propagation types and training types. In addition, they divided related applications into  structural scenarios, non-structural scenarios, and other scenarios. Zhang et al.\cite{zhang2019graphNeural} introduced GNNs on small graphs and giant graphs respectively. Quan et al. \cite{quan2019brief} and Zhang et al. \cite{zhang2019graphconvolutional} focused on reviewing works in a specific branch of GNNs,  i.e. graph convolutional network (GCN). However, they seldom introduce GNNs works in traffic scenarios.  Wu et.al proposed \cite{wu2020comprehensive} the only survey spending a paragraph to describe GNNs in traffic domain, which is obviously not enough for anyone desiring to explore this field.

In summary, there still lacks a systematic and elaborated survey to explore the rapidly developed graph-based deep learning techniques in traffic domain recently. Our work aims to fill this gap and promote understanding of the new emerging techniques in transportation community.

\section{Problems, Research Directions and Challenges}
\label{sec:Problems}

%%%%%%%%%%%%%%%%%%%%% Figures %%%%%%%%%%%%%%%%%%
\begin{figure}[htb]
\centering
\includegraphics[width=0.45\textwidth, height=0.17\textheight]{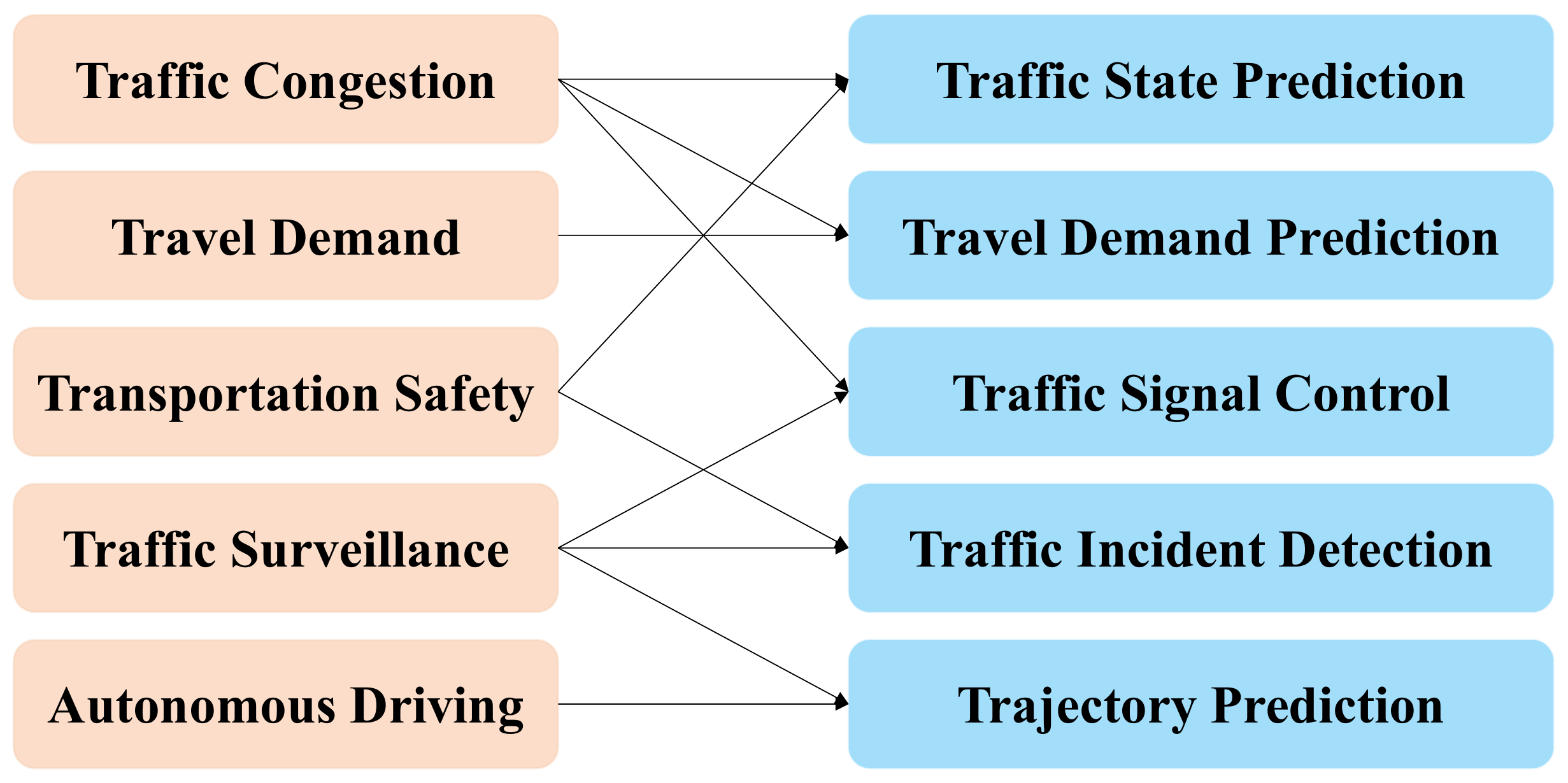}
\caption{Typical traffic problems and the corresponding research directions}
\label{fig:Class}
\end{figure}

In this section, we introduce background knowledge in traffic domain briefly, including some important traffic problems and the corresponding research directions (as shown in Figure \ref{fig:Class}), as well as common challenges and techniques under these problems. On one hand, we believe that such a concise but systematic introduction can help readers understand this domain quickly. On the other hand, our survey shows that existing works related with graph-based deep learning techniques only cover some research directions, which inspires successors to transfer similar techniques to remaining directions.

\subsection{Traffic Problems}
The goals the transportation community aims to achieve include relieving traffic congestion, satisfying travel demand, enhancing traffic management, ensuring transportation safety and realizing automatic driving. Each problem under the corresponding traffic goal can be partitioned into several research directions and each direction can serve more than one problem.

\subsubsection{\textbf{Traffic Congestion}}
Traffic congestion \cite{DBLP:conf/itsc/ChenLLW16} is one of the most important and urgent problems in modern cities in terms of significant time loss, air pollution and energy waste. The congestion can be solved by increasing the traffic efficiency \cite{DBLP:journals/tcps/YanS20}, \cite{DBLP:journals/ton/YanSC18}, alleviating the traffic congestion on road network \cite{ma2015large}, \cite{sun2017dxnat}, \cite{DBLP:conf/iotdi/YanSC17}, controlling the road conditions by traffic state prediction\cite{weidual19},\cite{chen2019multi}, optimizing vehicle flow by controlling traffic signals \cite{DBLP:journals/tits/CaoJZG17},\cite{DBLP:journals/tits/QiZL18}, optimizing passenger flow by predicting passenger demand in public transportation systems \cite{9207049}.

\subsubsection{\textbf{Travel Demand}}
The travel demand prediction refers to the demand of traffic services, such as taxi, bike, metro and bus in a crowd perspective. With the emerging of online ride-hailing platforms (e.g. Uber, DiDi) and rapid development of public transportation systems (e.g. metro system and bus system), travel demand prediction has become more and more important for transport authorities, business sectors and individuals. For related authorities, it can help to better allocate resources, e.g. increase metro frequency at rush hours, add more buses to service hotspots. For business sector, it enables them to better manage taxi-hiring \cite{rodrigues2019combining}, carpooling \cite{wang2017deepsd}, bike-sharing services \cite{DBLP:conf/gis/ChaiWY18},\cite{lin2018predicting}, and maximize their revenues. For individuals, it encourages users to consider various forms of transportation to decrease their commuting time and improve travel experience.

\subsubsection{\textbf{Transportation Safety}}
Transportation safety is an indispensable part of public safety. Traffic accidents can not only cause damage to victims, vehicles and road infrastructures, but also lead to traffic congestion and reduce efficiency of road network. Therefore, monitoring the traffic accidents is essential to avoid property loss and save life. Many researchers focus on directions such as detecting traffic incidents \cite{DBLP:conf/icde/HanGCWLS20}, predicting traffic accidents from social media data \cite{zhang2018deep}, predicting the injury severity of traffic accidents \cite{sameen2017severity}, \cite{alkheder2017severity}, predicting prevention of accidents \cite{DBLP:journals/tits/KashevnikLG20}, \cite{hanninen2014bayesian}, \cite{DBLP:journals/sensors/JoLKKK19}.

\subsubsection{\textbf{Traffic Surveillance}}
Nowadays, surveillance cameras have been widely deployed in city roads, generating numerous images and videos \cite{chen2019survey}. Such development has enhanced traffic surveillance, which includes traffic law enforcement, automatic toll collection \cite{qiu2019dynamic} and traffic monitoring systems. The research directions of traffic surveillance include license plate detection\cite{DBLP:journals/tits/LiWS19}, automatic vehicle detection \cite{chen2014vehicle}, pedestrian detection \cite{zhang2018occluded}.

\subsubsection{\textbf{Autonomous Driving}}
Recently, automatic driving vehicle has become a hot spot of research in transportation domain. Many tasks are related with visual recognition. The research directions of autonomous driving include lane/vehicle detection \cite{tayara2017vehicle}, pedestrian detection \cite{li2017scale}, traffic sign detection \cite{DBLP:journals/tits/KamalHI20} and human/vehicle trajectory prediction \cite{DBLP:journals/corr/abs-2002-06241}.

\subsection{Research Directions}
Our survey of graph-based deep learning in traffic domain shows that existing works focus mainly on traffic state prediction, travel demand prediction, trajectory prediction. A few works focus on vehicle behavior classification \cite{DBLP:journals/corr/abs-2002-00786}, optimal dynamic electronic toll collection (DETC) scheme \cite{qiu2019dynamic}, path availability \cite{DBLP:conf/kdd/LiHCSWZP19}, traffic signal control \cite{DBLP:conf/itsc/NishiOHY18}. To our best knowledge, traffic incident detection and vehicle detection have not been explored based in a graph view yet.

\subsubsection{\textbf{Traffic State Prediction}}
Traffic state in literatures refers to traffic flow, traffic speed, travel time, traffic density and so on. Traffic Flow Prediction (TFP) \cite{DBLP:conf/icpr/ZhangJCXP18},\cite{DBLP:conf/aaai/GuoLFSW19}, Traffic Speed Prediction (TSP) \cite{DBLP:conf/mdm/GeLLZ19}, \cite{DBLP:journals/corr/abs-1903-00919}, Travel Time Prediction (TTP) \cite{DBLP:conf/icde/HuG0J19},\cite{wang2018will}, \cite{DBLP:journals/imwut/YanSLSSQZX18} are hot branches of traffic state prediction and have attracted intensive studies.

\subsubsection{\textbf{Travel Demand Prediction}}
Travel demand prediction aims to estimate the future number of users who require traffic services.  It can be categorized into two kinds, i.e. zone-level demand prediction and origin-destination travel demand prediction. The former one aims to predict the future travel demand in each region of a city, for example, to predict future taxi request in each area of a city \cite{DBLP:journals/corr/abs-1905-11395},\cite{DBLP:conf/ijcai/BaiYK0S19}, or to predict the station-level passenger demand in subway system \cite{DBLP:journals/corr/abs-2001-04889}, \cite{DBLP:journals/corr/abs-1912-05693}, \cite{DBLP:conf/bigdata2/YeZZXZY20}, \cite{9207049} or to predict the bike hiring demand in each region of a city \cite{DBLP:conf/gis/ChaiWY18},\cite{lin2018predicting}. The latter one aims to predict the number of travel demand from one region to another, which can provide richer information than the zone-level demand prediction and is a more challenging issue worth exploration. Up to now, there are only a few studies \cite{DBLP:conf/icde/ShiYGLZYLL20}, \cite{DBLP:conf/icde/Hu0GJX20}, \cite{DBLP:journals/tits/ChuLL20a} directed towards the origin-destination based travel demand prediction, which is a promising research direction. 

\subsubsection{\textbf{Traffic Signal Control}}
The traffic signal control aims to properly control the traffic lights so as to reduce vehicle staying time at the road intersections in the long run \cite{wang2019enhancing}. Traffic signal control \cite{DBLP:conf/itsc/NishiOHY18} can optimize the traffic flow and reduce traffic congestion and vehicle emission.

\subsubsection{\textbf{Traffic Incident Detection}}
Major incidents can cause fatal injuries to travelers and long delays on a road network. Therefore, understanding the main cause of incidents and the impact of incidents on a traffic network is crucial for a modern transportation management system \cite{DBLP:conf/icde/HanGCWLS20},\cite{sameen2017severity}, \cite{alkheder2017severity}.

\subsubsection{\textbf{Human/Vehicle Trajectory Prediction}}
Trajectory Prediction \cite{DBLP:journals/corr/abs-2002-06241}, \cite{8729387}, \cite{DBLP:journals/corr/abs-2005-12661} aims to forecast future positions of dynamic agents in a scene. Accurate human/vehicle trajectories prediction is of great importance for downstream tasks including autonomous driving and traffic surveillance \cite{DBLP:conf/cvpr/MohamedQEC20}. For instance, an accurate pedestrian trajectory prediction can help controller to control the vehicle ahead in a dangerous environment \cite{DBLP:journals/corr/abs-2003-13924}. It can also enable transportation surveillance system to identify suspicious activities \cite{DBLP:conf/nips/KosarajuSM0RS19}.

\subsection{Challenges and Techniques Overview}
%%%%%%%%%%%%%%%%%%%%% Figures %%%%%%%%%%%%%%%%%%
\begin{figure}[htb]
\centering
\includegraphics[width=0.35\textwidth, height=0.2\textheight]{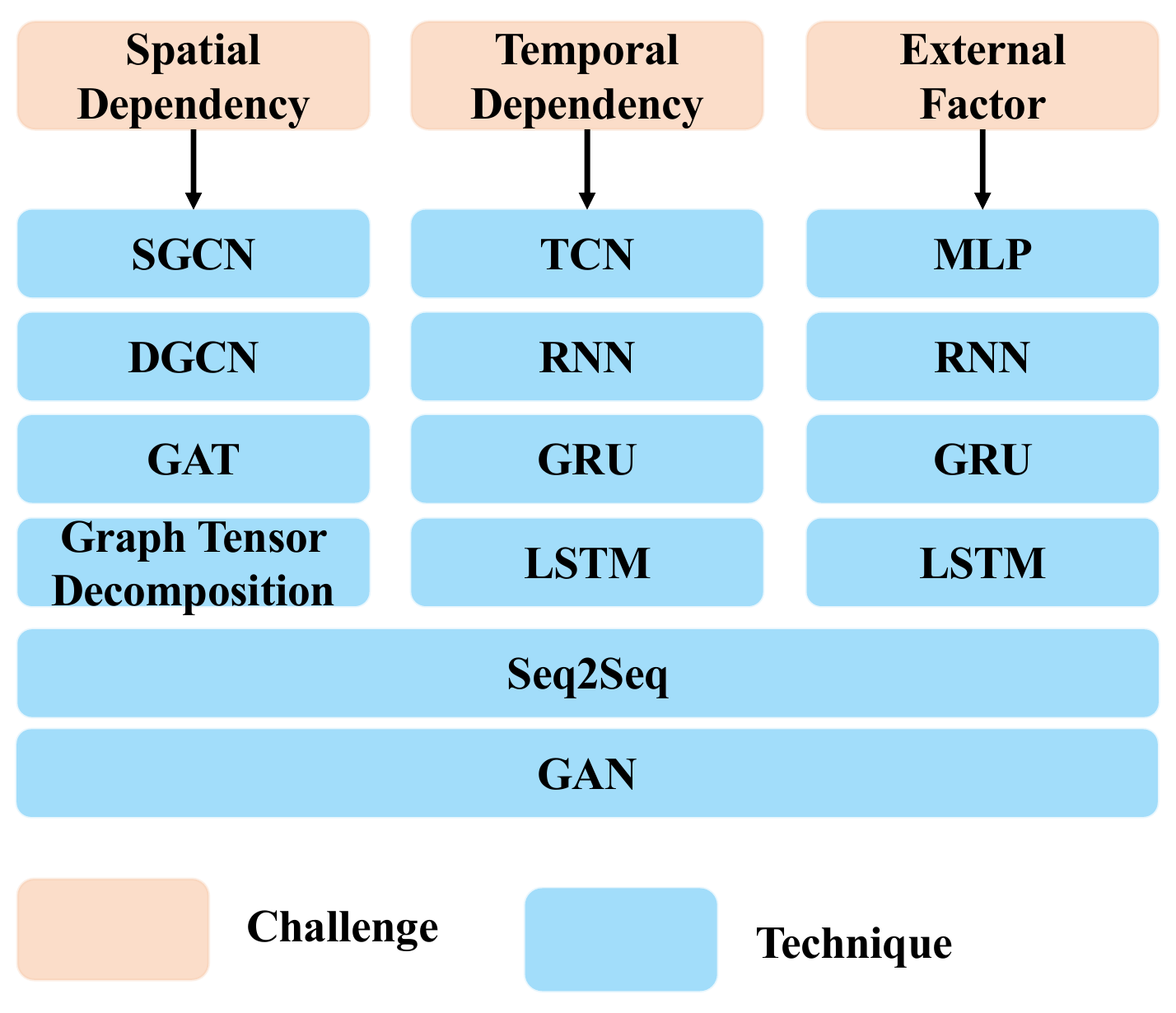}
\caption{Traffic challenges and the corresponding deep learning techniques. SGCN refers spectral graph convolution network, DGCN refers diffusion graph convolution network, GAT refers graph attention network, TCN refers temporal convolution network, RNN refers recurrent neural network, GRU refers gated recurrent unit, LSTM refers long short term memory network, MLP refers multi-layer perceptron, Seq2Seq refers sequence to sequence model, GAN refers generative adversarial network.}
\label{fig:Techniques}
\end{figure}

Although traffic problems and the related research directions are different, most of them share the same challenges, e.g. spatial dependency, temporal dependency, external factors. 

\subsubsection{\textbf{Spatiotemporal Dependency}}
There are complex spatiotemporal dependency in traffic data which can affect the prediction in traffic tasks. For instance, to predict a traffic congestion in a region, its previous traffic conditions and the traffic conditions of its surrounding regions are important factors for prediction\cite{DBLP:conf/itsc/ChenLLW16},\cite{ma2015large},\cite{sun2017dxnat}. 
In vehicle trajectory prediction, the stochastic behaviors of surrounding vehicles and the historical information of self-trajectory influence the prediction performance \cite{DBLP:journals/corr/abs-2003-11973}. 
When it comes to predict the ride-hailing demand in a region, its previous orders as well as orders in other regions with similar functionality are critical for prediction\cite{geng2019spatiotemporal}. 
To predict the traffic signal, the geometric features of multiple intersections are taken into consideration, as well as the previous traffic flow around \cite{DBLP:conf/itsc/NishiOHY18}.

\subsubsection{\textbf{External Factors}}
Except the spatiotemporal data, some types of data play an important role in traffic tasks, referred as external factors, such as holidays, weather conditions (e.g. rainfall, temperature, air quality),  extreme events \cite{laptev2017time} and traffic incidents (e.g. incident time, incident type) \cite{DBLP:journals/corr/abs-1912-01242}. The influence of external factors on traffic conditions can be observed in daily life. A rainstorm is likely to affect the traffic volume. A large-scale concert or football match results in traffic congregation, affecting traffic conditions around.

To tackle challenges above, various deep learning techniques have been proposed. In this paper, we focus on graph-based deep learning architectures in traffic domain. Among these graph-based deep learning frameworks, graph neural networks (GNNs) are usually employed to model the spatial dependency in traffic network. Recurrent neural networks (RNNs) and  temporal convolution network (TCN) are generally adopted to model the temporal dependency in traffic data. RNNs and Multi-layer Perceptrons (MLPs) are typically employed to process external factors. Sequence to Sequence (Seq2Seq) model is usually utilized to make multi-step traffic prediction. These techniques along with other tricks (e.g. gated mechanism, attention mechanism) are combined organically to improve the prediction accuracy.

In this paper, we aim to provide readers guidance about how to build a graph-based deep learning architecture and we have investigated enormous existing traffic works adopting graph-based deep learning solutions. In the following sections, we first introduce a common way to formulate the traffic problem and give detailed guidelines to build traffic graphs from various kinds of traffic data. Then we clarify the correlations between challenges and techniques (as shown in Figure \ref{fig:Techniques}) in two perspectives, i.e. the techniques perspective and the challenges perspective. In the perspective of techniques, we introduce several common techniques and interpret the way they tackle challenges in traffic tasks. In the perspective of challenges, we elaborate each challenge and summarize the techniques which can tackle this challenge. In a word, we hope to provide insights into solving traffic challenges with various deep learning techniques based on a graph view.

\section{Problem Formulation and Graph Construction}
\label{sec:GraphConstruction}
Among the graph-based deep learning traffic literatures we investigate, the majority of tasks (more than 80\%) belong to spatiotemporal forecasting problems, especially traffic state prediction and travel demand prediction. In this section, we first list commonly used notations. Then we summarize a general formulation of graph-based spatiotemporal prediction in traffic domain. We provide details to construct graphs from various traffic datasets. We also discuss multiple definitions of adjacency matrix, which represents the topology of graph-based traffic network and is the key element of graph-based solution.
%%%%%%%%%%%%%%%%%%%%%%%%% Table %%%%%%%%%%%%%%%%%%
\begin{table}[htb]
\caption{Notations In This Paper}
\tiny
\begin{tabular}{p{55pt}|p{175pt}}
\toprule
\multicolumn{2}{p{210pt}}{\textbf{Graph related elements}}\\
\midrule
$\mathbf{G}$  &Graph\\\hline
$\mathbf{E}$ &Edges of graph $\mathbf{G}$\\\hline
$\mathbf{V} $ &Vertices of graph $\mathbf{G}$\\\hline
$\mathbf{A}\in \mathbb{R}^{\mathbf{N} \times \mathbf{N}}$ &Adjacency matrix of graph $\mathbf{G}$ \\\hline
$\mathbf{A}^{T}\in \mathbb{R}^{\mathbf{N} \times \mathbf{N}}$ &The transpose matrix of $\mathbf{A}$ \\\hline
$\mathbf{\tilde{A}}\in \mathbb{R}^{\mathbf{N} \times \mathbf{N}}$ &Equal to $\mathbf{A}+\mathbf{I_{N}}$, a self-looped $\mathbf{A}$  \\\hline
$\mathbf{D}\in \mathbb{R}^{\mathbf{N} \times \mathbf{N}}$ &The degree matrix of adjacency matrix $\mathbf{A} $  \\\hline
$\mathbf{D_{I}}\in \mathbb{R}^{\mathbf{N} \times \mathbf{N}}$ &The in-degree matrix of adjacency matrix $\mathbf{A} $ \\\hline
$\mathbf{D_{O}}\in \mathbb{R}^{\mathbf{N} \times \mathbf{N}}$ &The out-degree matrix of adjacency matrix $\mathbf{A} $ \\\hline
$\mathbf{L}\in \mathbb{R}^{\mathbf{N} \times \mathbf{N}}$ &Laplacian matrix of graph $\mathbf{G}$ \\\hline
$\mathbf{U}\in \mathbb{R}^{\mathbf{N} \times \mathbf{N}}$  &The eigenvectors matrix of $\mathbf{L}$ \\\hline
$\mathbf{\Lambda}\in \mathbb{R}^{\mathbf{N} \times \mathbf{N}}$ & The diagonal eigenvalues matrix of $\mathbf{L}$ \\\hline
$\boldsymbol{\lambda}_{max}$ &The max eigenvalue of $\mathbf{L}$\\\hline
$\mathbf{I_{N}}\in \mathbb{R}^{\mathbf{N} \times \mathbf{N}}$  &An identity matrix \\
\midrule
\multicolumn{2}{p{210pt}}{\textbf{Hyper parameters}}\\
\midrule
$\mathbf{N}$  &The number of nodes in graph $\mathbf{G}$\\\hline
$\mathbf{F_{I}}$  &The number of input features\\\hline
$\mathbf{F_{H}}$  &The number of hidden features\\\hline
$\mathbf{F_{O}}$  &The number of output features\\\hline
$\mathbf{P}$  &The number of past time slices \\\hline
$\mathbf{Q}$  &The number of future time slices \\\hline
$\mathbf{d}$ &The dilation rate\\
\midrule
\multicolumn{2}{p{210pt}}{\textbf{Trainable parameters}}\\
\midrule
$W,b,\theta,\phi$  &The trainable parameters\\\hline
$\Theta$ &The kernel\\
\midrule
\multicolumn{2}{p{210pt}}{\textbf{Activation functions}}\\
\midrule
$\boldsymbol{\rho}(\boldsymbol{\cdot})$ &The activation function, e.g. tanh, sigmoid, ReLU\\\hline
$\boldsymbol{\sigma}(\boldsymbol{\cdot})\in[0,1]$ &The sigmoid function\\\hline
$\boldsymbol{tanh}(\boldsymbol{\cdot})\in[-1,1]$ &The hyperbolic tangent function\\\hline
$\boldsymbol{ReLU}(\boldsymbol{\cdot})\in[0,x]$ &The ReLU function\\
\midrule
\multicolumn{2}{p{210pt}}{\textbf{Operations}}\\
\midrule
$\boldsymbol{*_{\mathcal{G}}}$  &The convolution operator on graph\\\hline
$\boldsymbol{\odot}$ &Element-wise multiplication\\\hline
$\boldsymbol{\cdot}$ &Matrix multiplication\\
\midrule
\multicolumn{2}{p{210pt}}{\textbf{Spatial variables}}\\
\midrule
$X\in \mathbb{R}^{\mathbf{N} \times \mathbf{F_{I}}}$  &An input graph composed of $\mathbf{N}$ nodes with $\mathbf{F_{I}}$ features\\\hline
$X_{j}\in \mathbb{R}^{\mathbf{N}}$  &The $j^{th}$ feature of an input graph \\\hline
$X^{i}\in \mathbb{R}^{\mathbf{F_{I}}}$   &Node $i$ in an input graph\\\hline
$x \in \mathbb{R}^{\mathbf{N}}$  &A simply input graph \\\hline
$Y\in \mathbb{R}^{\mathbf{N} \times \mathbf{F_{O}}}$  &An output graph composed of $\mathbf{N}$ nodes with $\mathbf{\mathbf{F_{O}}}$ features\\\hline
$Y_{j}\in \mathbb{R}^{\mathbf{N}}$  &The $j^{th}$ feature of an output graph \\\hline
$Y^{i}\in \mathbb{R}^{\mathbf{F_{O}}}$  & Node $i$ in an output graph\\\hline
$y \in \mathbb{R}^{\mathbf{N}}$  &A simply output graph \\
\midrule
\multicolumn{2}{p{210pt}}{\textbf{Temporal variables}}\\
\midrule
$\mathbf{X} \in \mathbb{R}^{\mathbf{P}\times \mathbf{F_{I}}}$ &A sequential input with $\mathbf{F_{I}}$ features over $\mathbf{P}$ time slices\\\hline
$\mathbf{X}_{t} \in \mathbb{R}^{\mathbf{F_{I}}}$ &The element of sequential input at time $t$\\\hline
$\mathbf{x} \in \mathbb{R}^{\mathbf{P}}$ &A simply sequential input over $\mathbf{P}$ time slices\\\hline
$\mathbf{x}_{t} \in \mathbb{R}$ &The element of simply sequential input at time $t$\\\hline
$\mathbf{H}_{t} \in \mathbb{R}^{\mathbf{F_{H}}}$ &A hidden state with $\mathbf{\mathbf{F}_{H}}$ features at time $t$\\\hline
$\mathbf{Y} \in \mathbb{R}^{\mathbf{P}\times \mathbf{F_{O}}}$ &A sequential output with $\mathbf{F_{O}}$ features over $\mathbf{P}$ time slices\\\hline
$\mathbf{Y}_{t} \in \mathbb{R}^{\mathbf{F_{O}}}$ &The element of sequential output at time $t$\\\hline
$\mathbf{y} \in \mathbb{R}^{\mathbf{P}}$ &A simply sequential output over $\mathbf{P}$ time slices\\\hline
$\mathbf{y}_{t} \in \mathbb{R}$ &The element of simply sequential output at time $t$\\
\midrule
\multicolumn{2}{p{210pt}}{\textbf{Spatiotemporal variables}}\\
\midrule
$\mathcal{X}\in \mathbb{R}^{\mathbf{P} \times \mathbf{N} \times \mathbf{F_{I}}}$ &A series of  input graphs composed of  $\mathbf{N}$ nodes with $\mathbf{F_{I}}$ features over $\mathbf{P}$ time slices\\\hline
$\mathcal{X}_{t}\in \mathbb{R}^{\mathbf{N} \times \mathbf{F_{I}}}$  &An input graph at  time $t$\\\hline
$\mathcal{X}^{i}_{t}\in \mathbb{R}^{\mathbf{F_{I}}}$    &node $i$ in an input graph at time $t$\\\hline
$\mathcal{X}_{t,j}\in \mathbb{R}^{\mathbf{N}}$  &the $j^{th}$ feature of an input graph  at time $t$\\\hline
$\mathcal{X}^{i}_{t,j}\in \mathbb{R}$  &the $j^{th}$ feature of node $i$ in an input graph at time $t$\\\hline
$\mathcal{Y}\in \mathbb{R}^{\mathbf{P} \times \mathbf{N} \times \mathbf{F_{O}}}$ &A series of  output graphs composed of  $\mathbf{N}$ nodes with $\mathbf{F_{O}}$ features over $\mathbf{P}$ time slices\\\hline
$\mathcal{Y}_{t}\in \mathbb{R}^{\mathbf{N} \times \mathbf{F_{O}}}$  &An output graph at  time $t$\\\hline
$\mathcal{Y}^{i}_{t}\in \mathbb{R}^{\mathbf{F_{O}}}$    &node $i$ in an output graph at time $t$\\\hline
$\mathcal{Y}_{t,j}\in \mathbb{R}^{\mathbf{N}}$  &the $j^{th}$ feature of an output graph  at time $t$\\\hline
$\mathcal{Y}^{i}_{t,j}\in \mathbb{R}$  &the $j^{th}$ feature of node $i$ in an output graph at time $t$\\
\bottomrule
\end{tabular}
\label{tab:notation}
\end{table}

\subsection{Notations}
In this section, we have denoted some commonly used notations, including graph related elements, variables, parameters (hyper or trainable), activation functions, and operations. The variables are comprised of input variables \{$x$, $X$, $\mathbf{x}$, $\mathbf{X}$, $\mathcal{X}$\} and output variables \{$y$, $Y$, $\mathbf{y}$, $\mathbf{Y}$, $\mathcal{Y}$\}. 
These variables can divided into spatial variables, temporal variables, spatiotemporal variables. The spatial variables are only related with spatial attributes and the temporal variables are only related with temporal attributes. The spatiotemporal variables are related with both spatial and temporal attributes.

\subsection{Graph-based Spatio-Temporal Forecasting}
To our best knowledge, most existing graph-based deep learning traffic works can be categorized into spatial-temporal forecasting due to that most traffic datasets have both spatial attributes and temporal attributes. They formalize their prediction problems in a very similar way despite different mathematical notations and representations. We summarize their works to provide a general formulation for graph-based spatial-temporal problems in traffic domain.

The traffic network is represented as a graph  $\mathbf{G} = (\mathbf{V},\mathbf{E},\mathbf{A})$, which can be weighted \cite{DBLP:conf/ijcai/YuYZ18},\cite{DBLP:conf/icde/HuG0J19},\cite{DBLP:conf/icpr/ZhangJCXP18} or unweighted \cite{DBLP:conf/kdd/LiHCSWZP19},\cite{zhaoTGCN19},\cite{cuitraffic19}, directed \cite{DBLP:conf/kdd/LiHCSWZP19},\cite{DBLP:journals/tits/YuG19},\cite{DBLP:conf/trustcom/HuangWYC19} or undirected \cite{DBLP:conf/aaai/GuoLFSW19},\cite{DBLP:conf/uic/LiPLXDMWB18}, depending on specific tasks. $\mathbf{V}$ is a set of nodes and $|\mathbf{V}|= \mathbf{N}$ refers $\mathbf{N}$ nodes in the graph. Each node represents a traffic object, which can be a sensor \cite{DBLP:conf/mdm/GeLLZ19},\cite{DBLP:conf/aaai/GuoLFSW19},\cite{DBLP:journals/corr/abs-1911-08415}, a road segment \cite{DBLP:conf/ijcai/YuYZ18},\cite{DBLP:conf/aaai/Diao0ZLXH19},\cite{DBLP:conf/ijcnn/ZhangWCC19}, a road intersection \cite{DBLP:conf/icde/HuG0J19},\cite{DBLP:journals/tits/YuG19}, \cite{DBLP:conf/icpr/ZhangJCXP18}. $\mathbf{E}$ is a set of edges referring the connectivity between nodes. 

$\mathbf{A}=(\mathbf{a}_{ij})_{\mathbf{N}\times \mathbf{N}} \in \mathbb{R}^{\mathbf{N} \times \mathbf{N}}$ is the adjacency matrix containing the topology information of the traffic network, which is valuable for traffic prediction. The entry $\mathbf{a}_{ij}$ in matrix $\mathbf{A}$ represents the node proximity and is different in various applications. It can be a binary value $0$ or $1$ \cite{DBLP:conf/aaai/GuoLFSW19},\cite{zhaoTGCN19},\cite{cuitraffic19}. Specifically, $0$ indicates no edge between node $i$ and node $j$  while $1$ indicates an edge between these two nodes. It can also be a float value representing some kind of relationship between nodes \cite{DBLP:conf/ijcai/YuYZ18},\cite{DBLP:conf/kdd/WangYCW0019}, e.g. the road distance between two sensors \cite{DBLP:conf/mdm/GeLLZ19},\cite{DBLP:conf/ijcai/WuPLJZ19},\cite{DBLP:conf/trustcom/HuangWYC19}. 

$\mathcal{X}_{t} = [\mathcal{X}_{t}^{1},\cdots,\mathcal{X}_{t}^{i},\cdots,\mathcal{X}_{t}^{\mathbf{N}}]\in \mathbb{R}^{\mathbf{N} \times \mathbf{F_{I}}}$ is a feature matrix of the whole graph at time $t$.  $\mathcal{X}^{i}_{t}\in \mathbb{R}^{\mathbf{F_{I}}}$ represents node $i$ with $\mathbf{F_{I}}$ features at time $t$. The features are usually traffic indicators, such as traffic flow \cite{DBLP:conf/uic/LiPLXDMWB18},\cite{DBLP:conf/trustcom/HuangWYC19}, traffic speed \cite{DBLP:conf/mdm/GeLLZ19},\cite{DBLP:conf/aaai/Diao0ZLXH19},\cite{DBLP:journals/tits/YuG19}, or rail-hail orders \cite{geng2019spatiotemporal},\cite{DBLP:conf/kdd/WangYCW0019}, passenger flow \cite{DBLP:journals/corr/abs-2001-04889},\cite{DBLP:journals/corr/abs-1912-05693}. Usually, continuous indicators are normalized during data preprocessing phase.

\begin{figure}[htb]
\centering
\includegraphics[width=0.4\textwidth, height=0.15\textheight]{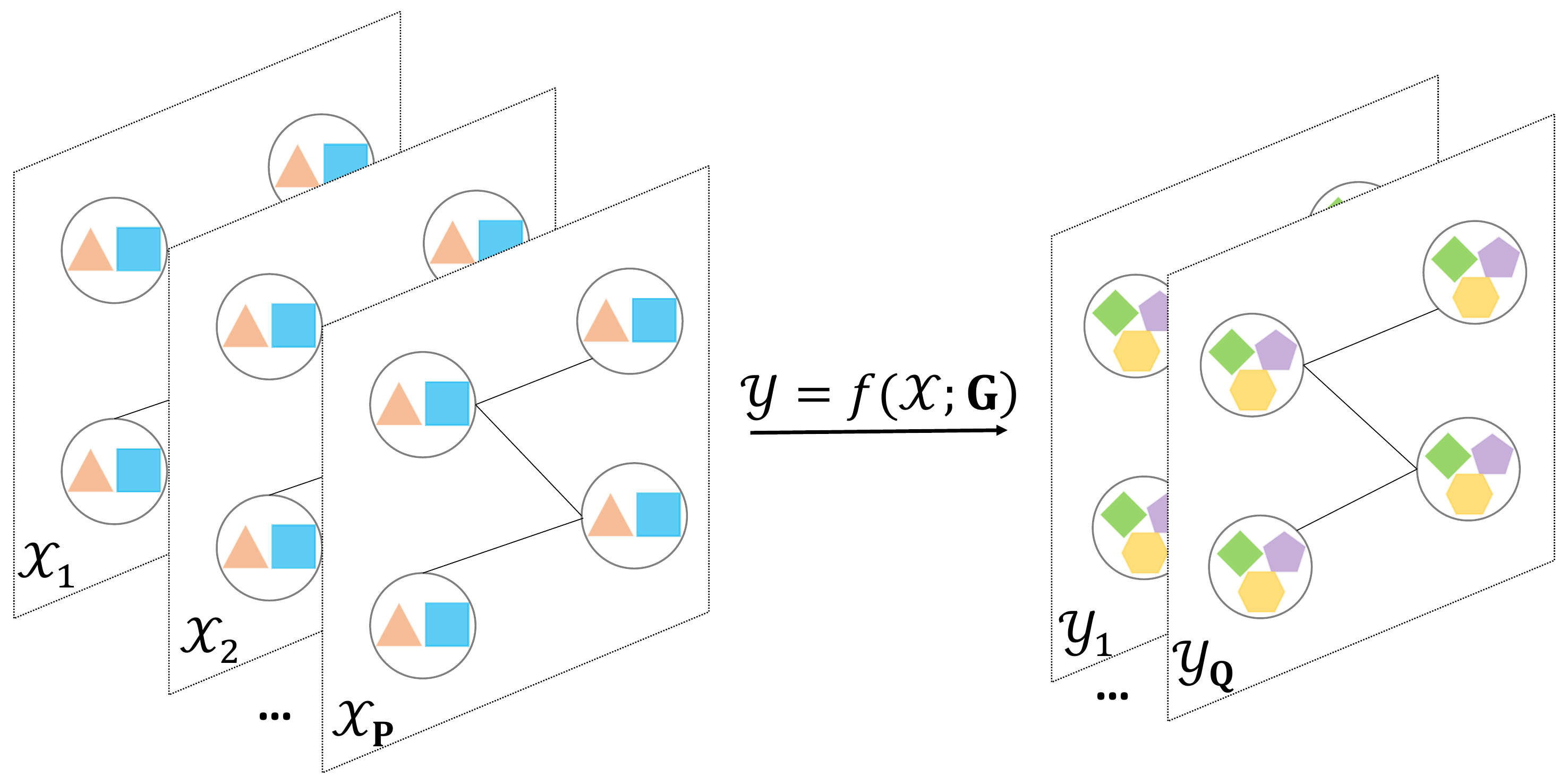}
\caption{The graph-based spatial-temporal problem formulation in traffic domain}
\label{fig:Formulation}
\end{figure}

Given historical indicators of the whole traffic network over past $\mathbf{P}$ time slices, denoted as $\mathcal{X} = [\mathcal{X}_{1},\cdots,\mathcal{X}_{t},\cdots,\mathcal{X}_{\mathbf{P}}]\in\mathbb{R}^{\mathbf{P} \times \mathbf{N} \times \mathbf{F_{I}}}$, the spatial-temporal forecasting problem in traffic domain aims to predict the future traffic indicators over the next $\mathbf{Q}$ time slices, denoted as $\mathcal{Y} = [\mathcal{Y}_{1},\cdots,\mathcal{Y}_{t},\cdots,\mathcal{Y}_{\mathbf{Q}}]\in\mathbb{R}^{\mathbf{Q} \times \mathbf{N} \times \mathbf{F_{O}}}$, where $\mathcal{Y}_{t}\in \mathbb{R}^{\mathbf{N} \times \mathbf{F_{O}}}$ represents output graph with $\mathbf{F_{O}}$ features at time $t$. The problem (as shown in Figure \ref{fig:Formulation}) can be formulated as follows:

\begin{footnotesize}\begin{equation}
\mathcal{Y} = f(\mathcal{X}; \mathbf{G})
\end{equation}\end{footnotesize}
%%%%%%%%%%%%%%%%%%%%% Figures %%%%%%%%%%%%%%%%%%

Some works predict multiple traffic indicators in the future (i.e. $\mathbf{F_{O}}>1$)  while other works predict one traffic indicator (i.e. $\mathbf{F_{O}}=1$), such as traffic speed \cite{DBLP:conf/aaai/Diao0ZLXH19}, \cite{DBLP:journals/tits/YuG19}, rail-hide orders \cite{geng2019spatiotemporal},\cite{DBLP:conf/kdd/WangYCW0019}. 
Some works only consider one-step prediction \cite{DBLP:journals/corr/abs-1910-09103},\cite{DBLP:journals/corr/abs-1905-11395},\cite{qiu2019dynamic}, i.e. forecasting traffic conditions in the next time step and $\mathbf{Q}=1$. But models designed for one-step prediction can't be directly applied to predict multiple steps, because they are optimized by reducing error during the training stage for the next-step instead of the subsequent time steps \cite{DBLP:conf/ijcai/BaiYK0S19}. Many works focus on multi-step forecasting  (i.e. $\mathbf{Q}>1$)  \cite{kanglearning19},\cite{chen2019multi},\cite{DBLP:journals/corr/abs-1903-05631}. According to our survey, there are mainly three kinds of techniques to generate a multi-step output, i.e. FC layer, Seq2Seq, dilation technique.  Fully connected (FC) layer is the simplest technique as being the output layer to obtain a desired output shape \cite{DBLP:conf/mdm/GeLLZ19}, \cite{DBLP:conf/aaai/GuoLFSW19}, \cite{guooptimized20}, \cite{zhaoTGCN19}, \cite{DBLP:journals/corr/abs-1912-01242}, \cite{DBLP:journals/access/ZhangYL19a}. Some works adopt the Sequence to Sequence (Seq2Seq) architecture with a RNNs-based decoder to generate output recursively through multiple steps \cite{DBLP:conf/iclr/LiYS018},\cite{DBLP:journals/corr/abs-1911-08415},\cite{DBLP:journals/corr/abs-1903-06261},\cite{kanglearning19},\cite{zhang2019multistep},\cite{DBLP:conf/trustcom/HuangWYC19}. Dilation technique is adopted  to get a desired output length \cite{DBLP:conf/ijcai/WuPLJZ19}, \cite{DBLP:journals/corr/abs-1903-05631}.
%%%%%%%%%%%%%%%%%%% Figure %%%%%%%%%%%%%%%%%%%%
\begin{figure*}[htb]
\centering
\includegraphics[width=0.7\textwidth, height=0.17\textheight]{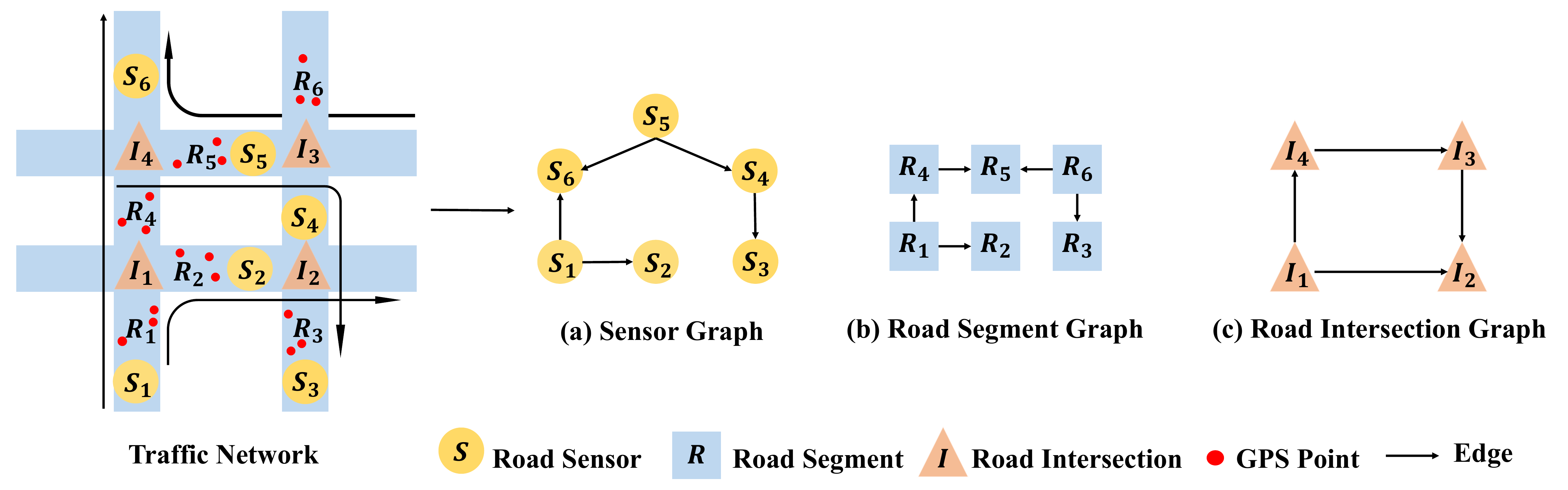}
\caption{Graph construction from various traffic datasets: a) In a sensor graph, sensor represents node and there is an edge between adjacent sensors on the same side of a road. b) In a road segment graph, road segment represents node and two connected segments have an edge. c) In a road intersection graph, road intersection represents node and there is an edge between two road intersections connected by a road segment. Most works consider the edge direction being the traffic flow direction\cite{DBLP:conf/mdm/GeLLZ19},\cite{DBLP:journals/corr/abs-1911-08415},\cite{DBLP:conf/kdd/LiHCSWZP19},\cite{DBLP:conf/trustcom/HuangWYC19},\cite{DBLP:conf/icpr/ZhangJCXP18},\cite{DBLP:conf/ijcai/FangZMXP19}, while some works ignore the direction and construct an undirected graph \cite{DBLP:conf/aaai/GuoLFSW19},\cite{DBLP:conf/ijcai/WuPLJZ19},\cite{cuitraffic19}\cite{DBLP:conf/ijcnn/ZhangWCC19},\cite{DBLP:journals/tits/YuG19}.}
\label{fig:road}
\end{figure*}
In addition, some works not only consider traffic indicators, but also take external factors (e.g. time attributes, weather) \cite{DBLP:conf/mdm/GeLLZ19},\cite{DBLP:conf/aaai/ChenLTZWWZ19},\cite{DBLP:journals/corr/abs-1912-01242},\cite{DBLP:conf/cikm/BaiYK0LY19} into consideration. Therefore, the problem formulation becomes: 

\begin{footnotesize}\begin{equation}\setlength{\abovedisplayskip}{0pt}\setlength{\belowdisplayskip}{0pt}
\mathcal{Y} = f(\mathcal{X},\mathcal{E}; \mathbf{G})
\end{equation}\end{footnotesize}where $\mathcal{E}$ is the external factors.
\subsection{Graph Construction from Traffic Datasets}
To model a traffic network as a graph is vital for any works that intend to utilize graph-based deep learning architectures to solve traffic problems. A traffic graph $\mathbf{G}$ for prediction is generally composed of four parts, i.e. nodes $\mathbf{V}$, node features (feature matrix $\mathcal{X}_{t}$), edges $\mathbf{E}$, edge weight $\mathbf{a}_{ij}$. Note that edges and edge weight can be represented by adjacency matrix $\mathbf{A}=(\mathbf{a}_{ij})_{\mathbf{N}\times \mathbf{N}}$. Nodes and node features can be constructed from traffic datasets. The construction of adjacency matrix not only depends on traffic datasets but also depends on the assumption of node relationship, which can be static or dynamic. We first introduce how to construct node and node features from various kinds of traffic datasets and then we give a systematic introduction to the popular adjacency matrices.

\subsubsection{\textbf{Nodes and Node Features  Construction}}
Many works are different in graph construction due to the different traffic datasets they collect. We divide these datasets into four categories according to the traffic infrastructures: data collected by the sensors deployed on road network \cite{DBLP:conf/mdm/GeLLZ19},\cite{DBLP:conf/aaai/GuoLFSW19},\cite{DBLP:journals/corr/abs-1903-00919}, vehicle GPS trajectories \cite{DBLP:conf/icpr/ZhangJCXP18},\cite{DBLP:conf/ijcai/FangZMXP19},\cite{DBLP:journals/tits/YuG19}, orders of rail-hailing system \cite{DBLP:conf/kdd/WangYCW0019},\cite{DBLP:conf/ijcai/BaiYK0S19},\cite{DBLP:conf/cikm/BaiYK0LY19}, transaction records of subway system \cite{DBLP:journals/corr/abs-2001-04889},\cite{DBLP:journals/corr/abs-1912-05693} or bus system \cite{DBLP:conf/ijcai/FangZMXP19}. For each category, we describe the datasets and explain the construction of nodes $\mathbf{V}$,  feature matrix $\mathcal{X}_{t}$.

\textbf{Sensors Datasets}
Traffic measurements (e.g. traffic speed) are generally collected during a short time interval by the sensors (e.g. loop detectors, probes) on a road network in metropolises like Beijing \cite{DBLP:conf/ijcai/YuYZ18}, California \cite{DBLP:journals/corr/abs-1903-00919}, Los  Angeles \cite{DBLP:conf/mdm/GeLLZ19}, New York \cite{DBLP:conf/aaai/Diao0ZLXH19}, Philadelphia \cite{guooptimized20}, Seattle \cite{cuitraffic19}, Xiamen \cite{DBLP:journals/corr/abs-1911-08415}, and Washington \cite{guooptimized20}. Sensor datasets are the most prevalent datasets in existing works, especially PEMS dataset from California. Generally, a road network contains traffic objects such as sensors, road segments. 

A sensor graph (as shown in Figure \ref{fig:road}) is constructed in \cite{DBLP:conf/mdm/GeLLZ19},\cite{DBLP:conf/aaai/GuoLFSW19},\cite{DBLP:conf/trustcom/HuangWYC19} where a sensor represents a node and features of this node are traffic measurements collected by its corresponding sensor.

A road segment graph (as shown in Figure \ref{fig:road}) is constructed in \cite{DBLP:conf/ijcai/YuYZ18},\cite{DBLP:conf/aaai/Diao0ZLXH19},\cite{guooptimized20} where a road segment represents a node and features of this node are average traffic measurements (e.g. traffic speed) recorded by all the sensors on its corresponding road segment.

\textbf{GPS Datasets}
GPS trajectories datasets are usually generated by numbers of taxis over some period of time in a city, e.g. Beijing \cite{DBLP:conf/icpr/ZhangJCXP18}, Chengdu \cite{DBLP:conf/icpr/ZhangJCXP18}, Shenzhen \cite{zhaoTGCN19}, Cologne \cite{DBLP:journals/tits/YuG19}, and Chicago \cite{DBLP:conf/ijcnn/ZhangWCC19}. Each taxi produces substantial GPS records with time, location, speed information every day. Every GPS record is fitted to its nearest road on the city road map. All roads are divided into multiple road segments through road intersections. 

A road segment graph (as shown in Figure \ref{fig:road}) is constructed in \cite{DBLP:conf/ijcnn/ZhangWCC19}, \cite{zhaoTGCN19} where a road segment represents a node and features of this node are average traffic measurements recorded by all the GPS points on its corresponding road segment.

A road intersection graph (as shown in Figure \ref{fig:road}) is constructed in \cite{DBLP:conf/icde/HuG0J19},\cite{DBLP:conf/icpr/ZhangJCXP18},\cite{DBLP:journals/tits/YuG19} where a road intersection represents a node and features of this node are sum-up of the traffic measurements through it. 

\begin{figure*}[htb]
\centering
\includegraphics[width=0.8\textwidth, height=0.15\textheight]{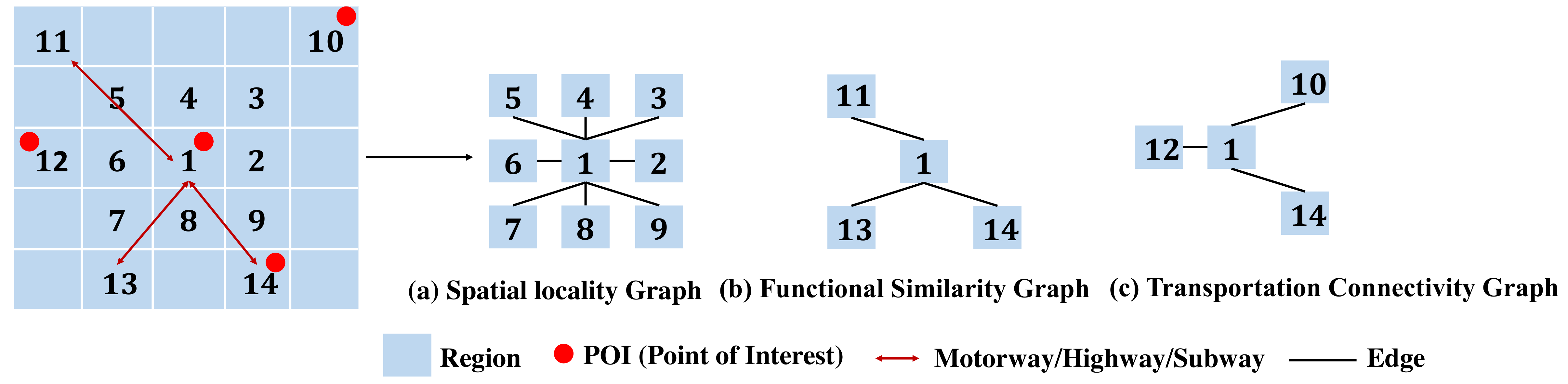}
\caption{Multi-relationships: a) A spatial locality graph: This graph is based on spatial proximity and it constructs edges between a region and its 8 adjacent regions in a 3 x 3 grid.
b) A functional similarity graph: This graph assumes that regions sharing similar functionality might have similar demand patterns. Edges are constructed between regions with similar surrounding POIs (Point of Interests).
c) A transportation connectivity graph: This graph assumes that regions which are geographically distant from the target region but conveniently reachable by transportation (e.g. motorway, highway or subway) have strong correlations with the target region. There should be edges between them.}
\label{fig:Multi-relationships}
\end{figure*}

\textbf{Rail-hailing Datasets}
These datasets record car/taxi/bicycle demand orders over a period of time in cities like Beijing \cite{geng2019spatiotemporal},\cite{DBLP:conf/kdd/WangYCW0019}, Chengdu \cite{DBLP:conf/kdd/WangYCW0019}, and Shanghai \cite{geng2019spatiotemporal}, Manhattan,  New York \cite{DBLP:conf/aaai/Diao0ZLXH19}. The target city with an OpenStreetMap is divided into equal-size grid-based regions (as shown in Figure \ref{fig:Multi-relationships}). Each region is defined as a node in a graph. The feature of each node is the number of orders in the corresponding region during a given interval. 

%%%%%%%%%%%%%%%%%%%%% Figures %%%%%%%%%%%%%%%%%%

\textbf{Transactions Datasets}
These datasets are collected by automatic fare collection (AFC) system deployed in public transit network, such as subway network and bus network.
A subway graph is constructed in \cite{DBLP:journals/corr/abs-2001-04889},\cite{DBLP:journals/corr/abs-1912-05693},\cite{DBLP:conf/ijcai/FangZMXP19}. Each station in the subway system is treated as a node. The features of a station usually contain the number of passengers departing at the station and the number of passengers arriving at the station during a given time interval based on transaction records collected by subwayAFC systems, which log when each passenger enters and leaves a metro system.

A bus graph is constructed in \cite{DBLP:conf/ijcai/FangZMXP19}. Each bus stop is treated as a node. The features of a bus stop usually contain the number of departing passengers at the station during a given time interval, but not the number of arriving passengers, since most bus AFC systems only log the boarding record of each passenger.

\subsubsection{\textbf{Adjacency Matrix Construction}}
The adjacency matrix $\mathbf{A}=(\mathbf{a}_{ij})_{\mathbf{N}\times \mathbf{N}} \in \mathbb{R}^{\mathbf{N} \times \mathbf{N}}$ is the key to capture spatial dependency which is valuable for prediction. Element $\mathbf{a}_{ij}$  (unweighted or weighted) represents heterogeneous pairwise relationship between nodes. However, there are different assumptions of node relationships in different traffic scenarios, based on which the adjacency matrix can be designed differently, e.g. fixed matrix, dynamic matrix, evolving matrix.

\textbf{Fixed Matrix}
Many works assume that the correlations between nodes are fixed and do not change over time. Therefore, a fixed matrix is designed and unchanged during the whole experiment. Researchers have designed various fixed adjacency matrices to capture various kinds of pre-defined correlations between nodes in a traffic graph, like function similarity and transportation connectivity \cite{geng2019spatiotemporal}, semantic connection \cite{DBLP:conf/kdd/WangYCW0019}, temporal similarity \cite{DBLP:journals/corr/abs-1903-00919}. Here, we introduce several popular adjacency matrices.

 Connection matrix measures the connectivity between nodes. The entry value in the matrix is defined as $1$ (connection) or $0$ (disconnection) \cite{DBLP:conf/aaai/GuoLFSW19},\cite{guooptimized20},\cite{zhaoTGCN19},\cite{cuitraffic19}.

Distance matrix measures the closeness between nodes in terms of geometrical distance. The entry value is defined as a function of distance between nodes \cite{DBLP:conf/cvpr/MohamedQEC20}. For example, some works \cite{DBLP:conf/icde/HuG0J19},\cite{DBLP:conf/icpr/ZhangJCXP18},\cite{DBLP:conf/ijcnn/ZhangWCC19},\cite{DBLP:conf/uic/LiPLXDMWB18},\cite{DBLP:conf/ijcai/BaiYK0S19},\cite{DBLP:journals/tits/YuG19} used threshold Gaussian Kernel to define $\mathbf{a}_{ij}$ as follows:

\begin{footnotesize}\begin{equation}\setlength{\abovedisplayskip}{2pt}\setlength{\belowdisplayskip}{2pt}
\mathbf{a}_{ij}=\left\{\begin{array}{l}\exp \left(-\frac{\mathbf{d}_{ij}^{2}}{\sigma^{2}}\right), i \neq j \text { and } \mathbf{d}_{ij} \geq \epsilon \\ 0 \quad, i = j \text { or } \mathbf{d}_{ij} < \epsilon \end{array}\right.
\end{equation}\end{footnotesize}where $\mathbf{d}_{ij}$ is the distance between node $i$ and node $j$. Hyper parameters $\sigma^{2}$ and $\epsilon$ are thresholds to control the distribution and sparsity of matrix $\mathbf{A}$.

Functional similarity matrix measures whether two nodes are similar in terms of functionality (e.g. both of them are business zones). The corresponding functional similarity graph is shown in Figure \ref{fig:Multi-relationships}. It assumes that regions sharing similar functionality might have similar demand patterns \cite{geng2019spatiotemporal}. Edges are constructed between regions with similar surrounding POIs (Point of Interests).

Transportation connectivity matrix measures the correlation between regions that are geographically distant but conveniently reachable by motorway, highway or subway. The corresponding transportation connectivity graph is shown in Figure \ref{fig:Multi-relationships}. There should be edges between them \cite{geng2019spatiotemporal}.

\textbf{Dynamic Matrix}
Some works argue that the pre-defined matrix does not necessarily reflect the true dependency among nodes due to the defective prior knowledge or incomplete data \cite{DBLP:conf/icde/HuG0J19}. A novel adaptive matrix is proposed and learned through data. Experiments in \cite{DBLP:conf/ijcai/WuPLJZ19},\cite{DBLP:conf/icde/HuG0J19},\cite{DBLP:conf/aaai/Diao0ZLXH19} have proven that adaptive matrix can precisely capture the hidden spatial dependency more precisely in some traffic tasks. 

\textbf{Evolving Matrix}
In some scenarios, the graph structure can evolve over time as some edges may become unavailable, like road congestion or closure, and become available again after alleviating congestion. An evolving topological structure \cite{DBLP:conf/kdd/LiHCSWZP19}, \cite{DBLP:conf/aaai/YanXL18} is incorporated into the model to capture such dynamic spatial change.

\section{Deep Learning Techniques Perspective}
\label{sec:Techniques}
%%%%%%%%%%%%%%%%%%%%%%%%%%%%%% Table %%%%%%%%%%%%%%%%%%%%
\begin{table*}[htb]
\caption{The decomposition of graph-based deep learning architectures investigated in this paper}
\centering
\scriptsize
\begin{tabular}{c|c|c|c|c}
\toprule
\textbf{Reference} &\textbf{Year}  &\textbf{Directions}  &\textbf{Models} &\textbf{Modules}\\
\midrule 
\cite{8729387}&2019&Human/Vehicle Trajectory Prediction&SAGCN&SGCN, TCN,  Attention\\\hline
\cite{DBLP:conf/nips/KosarajuSM0RS19}&2019&Human/Vehicle Trajectory Prediction&Social-BiGAT&GAT, LSTM, GAN\\\hline
\cite{DBLP:conf/cvpr/MohamedQEC20}&2020&Human/Vehicle Trajectory Prediction&Social-STGCNN&SGCN, TCN\\\hline
\cite{DBLP:journals/corr/abs-2002-06241}&2020&Human/Vehicle Trajectory Prediction&Social-WaGDAT&GAT, Seq2Seq, MLP\\\hline 
\cite{DBLP:journals/corr/abs-2003-11973}&2020&Human/Vehicle Trajectory Prediction&&SGCN, LSTM\\\hline
\cite{qiu2019dynamic}&2019&Optimal DETC Scheme&&SGCN\\\hline
\cite{DBLP:journals/corr/abs-2002-00786}&2020&Vehicle Behaviour Classification&MR-GCN&SGCN, LSTM\\\hline
\cite{DBLP:conf/itsc/NishiOHY18}&2018&Traffic Signal Control&&SGCN, Reinforcement Learning\\\hline
\cite{DBLP:conf/kdd/LiHCSWZP19}&2019&Path Availability&LRGCN-SAPE&SGCN, LSTM\\\hline
\cite{DBLP:conf/icde/HuG0J19}&2019&Travel Time Prediction&&SGCN\\\hline
\cite{DBLP:conf/icpr/ZhangJCXP18}&2018&Traffic Flow Prediction&KW-GCN&SGCN, LCN\\\hline
\cite{DBLP:conf/uic/LiPLXDMWB18}&2018&Traffic Flow Prediction&Graph-CNN&CNN, Graph Matrix\\\hline
\cite{wang2018dynamic}&2018&Traffic Flow Prediction&DST-GCNN&SGCN\\\hline
\cite{DBLP:conf/aaai/GuoLFSW19}&2019&Traffic Flow Prediction&&SGCN, CNN, Attention Mechanism\\\hline
\cite{DBLP:conf/ijcai/FangZMXP19}&2019&Traffic Flow Prediction&&SGCN, TCN, Residual\\\hline
\cite{DBLP:journals/corr/abs-1903-06261}&2019&Traffic Flow Prediction&GHCRNN&SGCN, GRU, Seq2Seq\\\hline
\cite{kanglearning19}&2019&Traffic Flow Prediction&STGSA&GAT, GRU, Seq2Seq\\\hline
\cite{DBLP:conf/trustcom/HuangWYC19}&2019&Traffic Flow Prediction&DCRNN-RIL&DGCN, GRU, Seq2Seq\\\hline
\cite{DBLP:journals/corr/abs-1903-07789}&2019&Traffic Flow Prediction&MVGCN&SGCN, FNN, Gate Mechanism, Residual\\\hline
\cite{DBLP:journals/cacie/00390R19}&2019&Traffic Flow Prediction&STGI- ResNet&SGCN, Residual\\\hline
\cite{DBLP:journals/corr/abs-1906-00560}&2020&Traffic Flow Prediction&FlowConvGRU&DGCN, GRU\\\hline
\cite{9207049}&2020&Traffic Flow Prediction&Multi-STGCnet&SGCN, LSTM\\\hline
\cite{DBLP:conf/uai/ZhangSXMKY18}&2018&Traffic Speed Prediction&&GAT, GRU, Gate Mechanism\\\hline
\cite{DBLP:conf/mdm/GeLLZ19}&2019&Traffic Speed Prediction&GTCN&SGCN, TCN, Residual\\\hline
\cite{DBLP:journals/corr/abs-1903-00919}&2019&Traffic Speed Prediction&3D-TGCN&SGCN, Gate Mechanism\\\hline
\cite{DBLP:journals/corr/abs-1912-01242}&2019&Traffic Speed Prediction&DIGC-Net&SGCN, LSTM\\\hline
\cite{DBLP:journals/corr/abs-1909-07105}&2019&Traffic Speed Prediction&MW-TGC&SGCN, LSTM\\\hline
\cite{zhang2019multistep}&2019&Traffic Speed Prediction&AGC-Seq2Seq&SGCN, GRU, Seq2Seq, Attention Mechanism\\\hline
\cite{DBLP:journals/tits/YuG19}&2019&Traffic Speed Prediction&GCGA&SGCN, GAN\\\hline
\cite{DBLP:journals/access/ZhangYL19a}&2019&Traffic Speed Prediction&ST-GAT&GAT, LSTM\\\hline
\cite{DBLP:conf/ijcai/YuYZ18}&2018&Traffic State Prediction&STGCN&SGCN, TCN, Gate Mechanism\\\hline
\cite{DBLP:conf/iclr/LiYS018}&2018&Traffic State Prediction&DCRNN&DGCN, GRU, Seq2Seq\\\hline
\cite{DBLP:conf/aaai/Diao0ZLXH19}&2019&Traffic State Prediction&&SGCN, CNN, Gate Mechanism\\\hline
\cite{DBLP:conf/aaai/ChenLTZWWZ19}&2019&Traffic State Prediction&MRes-RGNN&DGCN, GRU, Residual, Gate Mechanism\\\hline
\cite{DBLP:conf/ijcnn/ZhangWCC19}&2019&Traffic State Prediction&GCGAN&DGCN, LSTM, GAN, Seq2Seq, Attention Mechanism\\\hline
\cite{DBLP:conf/ijcai/WuPLJZ19}&2019&Traffic State Prediction&Graph WaveNet&DGCN, TCN, Residual, Gate Mechanism\\\hline
\cite{zhaoTGCN19}&2019&Traffic State Prediction&T-GCN&SGCN, GRU\\\hline
\cite{cuitraffic19}&2019&Traffic State Prediction&TGC-LSTM&SGCN, LSTM\\\hline
\cite{weidual19}&2019&Traffic State Prediction&DualGraph&Seq2Seq, MLP, Graph Matirx\\\hline
\cite{DBLP:journals/corr/abs-1903-05631}&2019&Traffic State Prediction&ST-UNet&SGCN, GRU\\\hline
\cite{DBLP:journals/corr/abs-1911-08415}&2020&Traffic State Prediction&GMAN&GAT, Gate Mechanism, Seq2Seq, Attention Mechanism\\\hline
\cite{guooptimized20}&2020&Traffic State Prediction&OGCRNN&SGCN, GRU, Attention Mechanism\\\hline
\cite{chen2019multi}&2020&Traffic State Prediction&MRA-BGCN&SGCN, GRU, Seq2Seq, Attention Mechanism\\\hline
\cite{DBLP:conf/gis/ChaiWY18}&2018&Travel Demand-Bike&&SGCN, LSTM, Seq2Seq\\\hline
\cite{lin2018predicting}&2018&Travel Demand-Bike&GCNN-DDGF&SGCN, LSTM\\\hline
\cite{DBLP:journals/corr/abs-2001-04889}&2020&Travel Demand-Subway&PVCGN&SGCN, GRU, Seq2Seq, Attention Mechanism\\\hline
\cite{DBLP:journals/corr/abs-1912-05693}&2019&Travel Demand-Subway&WDGTC&Tensor Completion, Graph Matrix\\\hline
\cite{geng2019spatiotemporal}&2019&Travel Demand-Taxi&CGRNN&SGCN, RNN, Attention Mechanism, Gate Mechanism\\\hline
\cite{DBLP:conf/kdd/WangYCW0019}&2019&Travel Demand-Taxi&GEML&SGCN, LSTM\\\hline
\cite{DBLP:journals/corr/abs-1905-11395}&2019&Travel Demand-Taxi&MGCN&SGCN\\\hline
\cite{DBLP:conf/ijcai/BaiYK0S19}&2019&Travel Demand-Taxi&STG2Seq&SGCN, Seq2Seq, Attention Mechanism, Gate Mechanism, Residual\\\hline
\cite{DBLP:conf/cikm/BaiYK0LY19}&2019&Travel Demand-Taxi&&SGCN, LSTM, Seq2Seq\\\hline
\cite{DBLP:journals/corr/abs-1910-09103}&2019&Travel Demand-Taxi&ST-ED-RMGC&SGCN, LSTM, Seq2Seq, Residual\\
\bottomrule
\end{tabular}
\label{tab:architectures}
\end{table*}

We summarize the graph-based deep learning architectures in existing traffic literatures and find that most of them are composed of graph neural networks (GNNs) and other modules, such as recurrent neural networks (RNNs),  temporal convolution network (TCN), Sequence to Sequence (Seq2Seq) model, generative adversarial network (GAN) (as shown in Table \ref{tab:architectures}). It is the cooperation of GNNs and other deep learning techniques that achieves state-of-the-art performance in many traffic scenarios. This section aims to introduce the functionality, advantages, defects and variants of these techniques in traffic tasks, which can help participators understand how to utilize these deep learning techniques in traffic domain.
\subsection{GNNs}
%%%%%%%%%%%%%%%%%%%%% Figures %%%%%%%%%%%%%%%%%%
\begin{figure*}[htb]
\centering
\includegraphics[width=1.0\textwidth, height=0.16\textheight]{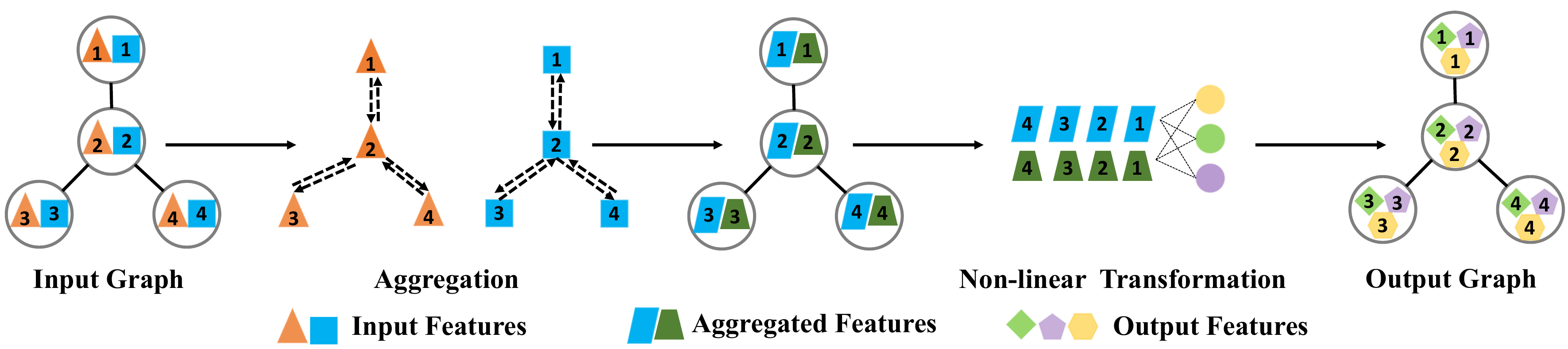}
\caption{The structure of Graph Neural Network is generally composed of two kind of layers: 1) Aggregation layer: In each feature dimension, the features of adjacent nodes are aggregated to the central node. Mathematically, the  output of aggregation layer is the product of  adjacency matrix and features matrix.
2) Non-linear transformation layer: All the aggregated features of each node are fed into the non-linear transformation layer to create higher level feature representation. All nodes share the same transformation kernel. $\{1,2,3,4\}$ are node indexes.}
\label{fig:GCN}
\end{figure*}

In the last couple of years, motivated by the huge success of deep learning approaches (e.g. CNNs, RNNs), there is an increasing interest in generalizing neural networks to arbitrarily structured graphs and such networks are classified as graph neural networks (GNNs).  In the early stage, the studies about GNNs can be categorized into recurrent graph neural networks
(RecGNNs) which are inspired by RNNs \cite{wu2020comprehensive}.  Subsequently, inspired by the huge success of CNNs, many works focus on extending the convolution of CNN on graph data and these works can be categorized into convolutional graph neural networks (ConvGNNs)  \cite{wu2020comprehensive}. There are also other branches of GNNs developed in recent years, e.g. graph auto-encoders (GAEs) \cite{DBLP:journals/nn/Majumdar18} and graph attention networks (GATs) \cite{DBLP:conf/iclr/VelickovicCCRLB18}.  According to our investigation, most traffic works focus on ConvGNNs and there are only a few studies \cite{DBLP:conf/uai/ZhangSXMKY18} employing other branches of GNNs up to now. Further, ConvGNNs can be divided into two main streams, i.e. the spectral-based approaches which develop graph convolutions based on the spectral theory and the spatial-based approaches which define graph convolutions based on spatial relations between nodes\cite{DBLP:conf/nips/HamiltonYL17}. Recently, many novel spatial-based convolutions have emerged, among which diffusion convolution is a popular spatial-based graph convolution which regards graph convolution as a diffusion process. 

According to our survey, most existing traffic works utilize either spectral graph convolution or diffusion graph convolution. There are also other novel convolutions \cite{DBLP:conf/icpr/ZhangJCXP18} but their applications in traffic domain are relatively few. Therefore, in this section, we focus on introducing spectral graph convolution (SGC) and diffusion graph convolution (DGC) in traffic domain. In this paper, we refer the graph neural network with spectral graph convolution as SGCN and that with diffusion graph convolution as DGCN. Note that SGC is for undirected graph while DGC can be applied in both directed graph and undirected graph. In addition, both SGC and DGC aim to generate new feature representations for each node  in a graph through feature aggregation and non-linear transformation (as shown in Figure \ref{fig:GCN}). 

\subsubsection{\textbf{Spectral Graph Convolution}}
In the spectral theory, a graph is represented by its corresponding normalized Laplacian matrix $\mathbf{L} = \mathbf{I_{N}}-\mathbf{D}^{-\frac{1}{2}} \mathbf{A} \mathbf{D}^{-\frac{1}{2}} \in\mathbb{R}^{\mathbf{N} \times \mathbf{N}}$. The real symmetric matrix $\mathbf{L}$ can be diagonalized via eigendecomposition as $\mathbf{L}=\mathbf{U} \mathbf{\Lambda} \mathbf{U}^{T}$ where $\mathbf{U}\in \mathbb{R}^{\mathbf{N} \times \mathbf{N}}$ is the eigenvectors matrix and $\mathbf{\Lambda}\in \mathbb{R}^{\mathbf{N} \times \mathbf{N}}$ is the diagonal eigenvalues matrix. Since $\mathbf{U}$ is also an orthogonal matrix, Shuman et al. \cite{DBLP:journals/spm/ShumanNFOV13} adopted it as a graph Fourier basis, defining graph Fourier transform of a graph signal $x \in \mathbb{R}^{\mathbf{N}}$ as $\hat{x}=\mathbf{U}^{T}x$, and its inverse as $x=\mathbf{U} \hat{x}$.

Bruna et al. \cite{DBLP:journals/corr/BrunaZSL13} tried to build an analogue of CNN convolution in spectral domain and defined the spectral convolution as $y=\Theta \boldsymbol{*_{\mathcal{G}}} x=\mathbf{U} \Theta \mathbf{U}^{T} x$, i.e. transforming $x$ into spectral domain, adjusting its amplitude by a diagonal kernel $\Theta=\operatorname{diag}(\theta_{0}, \ldots, \theta_{\mathbf{N}-1}) \in \mathbb{R}^{\mathbf{N} \times \mathbf{N}}$, and doing inverse Fourier transform to get the final result $y$ in spatial domain. Although such convolution is theoretically guaranteed, it is computationally expensive as multiplication with $\mathbf{U}$ is $\mathcal{O}(\mathbf{N}^2)$ and the eigendecomposition of $\mathbf{L}$ is intolerable for large scale graphs. In addition, it considers all nodes by the kernel $\Theta$ with $\mathbf{N}$ parameters and can't extract spatial localization. 

To avoid such limitations, Defferrard et al. \cite{DBLP:conf/nips/DefferrardBV16} localized the convolution and reduced its parameters by restricting the kernel $\Theta$ to be a polynomial of eigenvalues matrix $\mathbf{\Lambda}$ as $\Theta=\sum_{k=0}^{\mathbf{K}-1} \theta_{k} \mathbf{\Lambda}^{k}$ and $\mathbf{K}$ determines the maximum radius of the convolution from a central node. Thus, the convolution can be rewritten as $\Theta \boldsymbol{*_{\mathcal{G}}} x= \sum_{k=0}^{\mathbf{K}-1} \theta_{k} \mathbf{U}\mathbf{\Lambda}^{k} \mathbf{U}^{T} x=\sum_{k=0}^{\mathbf{K}-1} \theta_{k} \mathbf{L}^{k}x$. Further more, Defferrard et al. \cite{DBLP:conf/nips/DefferrardBV16} adopted the Chebyshev polynomials $ T_{k}(x)$ to approximate $\mathbf{L}^{k}$, resulting in $\Theta \boldsymbol{*_{\mathcal{G}}} x \approx \sum_{k=0}^{\mathbf{K}-1} \theta_{k}T_{k}(\tilde{\mathbf{L}})x$ with a rescaled $\tilde{\mathbf{L}}=\frac{2}{\boldsymbol{\lambda}_{\max}}\mathbf{L}-\mathbf{I_{N}}$ where $\boldsymbol{\lambda}_{\max}$ is the largest eigenvalue of $\mathbf{L}$ and $ T_{k}(x) =2xT_{k-1}(x)-T_{k-2}(x)$, $T_{0}(x)=1$, $T_{1}(x)=x$ \cite{hammond2011wavelets}. By recursively computing $T_{k}(x)$, the complexity of this $\mathbf{K}$-localized convolution can be reduced to $\mathcal{O}(\mathbf{K}|\mathbf{E}|)$ with $|\mathbf{E}|$ being the number of edges.

Based on \cite{DBLP:conf/nips/DefferrardBV16}, Kipf et al. \cite{DBLP:conf/iclr/KipfW17} simplified the spectral graph convolution by limiting $\mathbf{K}=2$ and with  $T_{0}(\tilde{\mathbf{L}})=1$, $T_{1}(\tilde{\mathbf{L}})=\tilde{\mathbf{L}}$. They got $\Theta \boldsymbol{*_{\mathcal{G}}} x\approx \theta_{0} T_{0}(\tilde{\mathbf{L}})x + \theta_{1}T_{1}(\tilde{\mathbf{L}})x = \theta_{0}x + \theta_{1}\tilde{\mathbf{L}}x$. Noticing that $\tilde{\mathbf{L}}=\frac{2}{\lambda_{\max}}\mathbf{L}-\mathbf{I_{N}}$, they set $\boldsymbol{\lambda}_{\max}=2$, resulting in  $\Theta \boldsymbol{*_{\mathcal{G}}} x \approx \theta_{0}x + \theta_{1}(\mathbf{L}-\mathbf{I_{N}})x$. For that $\mathbf{L} =\mathbf{I_{N}}-\mathbf{D}^{-\frac{1}{2}} \mathbf{A} \mathbf{D}^{-\frac{1}{2}}$ and $\mathbf{L}-\mathbf{I_{N}}=-\mathbf{D}^{-\frac{1}{2}} \mathbf{A} \mathbf{D}^{-\frac{1}{2}}$, they got $\Theta \boldsymbol{*_{\mathcal{G}}} x \approx \theta_{0}x - \theta_{1}(\mathbf{D}^{-\frac{1}{2}} \mathbf{A} \mathbf{D}^{-\frac{1}{2}})x$. Further, they reduced the number of parameters by setting $\theta=\theta_{0}=-\theta_{1}$ to address overfitting and got  $\Theta \boldsymbol{*_{\mathcal{G}}} x \approx\theta (\mathbf{I_{N}}+\mathbf{D}^{-\frac{1}{2}} \mathbf{A} \mathbf{D}^{-\frac{1}{2}})x $. They defined $\tilde{\mathbf{A}}=\mathbf{A}+\mathbf{I_{N}}$ and adopted a renormalization trick to get $y=\Theta \boldsymbol{\boldsymbol{*_{\mathcal{G}}}} x \approx \theta\tilde{\mathbf{D}}^{-\frac{1}{2}} \tilde{\mathbf{A}} \tilde{\mathbf{D}}^{-\frac{1}{2}}x$, where $\tilde{\mathbf{D}}$ is the degree matrix of $\tilde{\mathbf{A}}$. Finally, Kipf et al.\cite{DBLP:conf/iclr/KipfW17} proposed a spectral graph convolution layer as follows:
\begin{footnotesize}\begin{equation}\setlength{\abovedisplayskip}{2pt}\setlength{\belowdisplayskip}{2pt}
\begin{split}
Y_{j}&=\boldsymbol{\rho}(\Theta_{j} \boldsymbol{*_{\mathcal{G}}} X )=\boldsymbol{\rho}( \sum_{i=1}^{\mathbf{F_{I}}} \theta_{i, j}\tilde{\mathbf{D}}^{-\frac{1}{2}}\tilde{\mathbf{A}} \tilde{\mathbf{D}}^{-\frac{1}{2}}X_{i}), 1 \leq j \leq \mathbf{F_{O}}\\
Y&=\boldsymbol{\rho}(\tilde{\mathbf{D}}^{-\frac{1}{2}} \tilde{\mathbf{A}} \tilde{\mathbf{D}}^{-\frac{1}{2}} X W)
\end{split}\end{equation}\end{footnotesize}here, $X \in \mathbb{R}^{\mathbf{N} \times \mathbf{F_{I}}}$ is the layer input with $\mathbf{F_{I}}$ features, $X_{i}\in \mathbb{R}^{\mathbf{N}}$ is its $i^{th}$ feature. $Y \in \mathbb{R}^{\mathbf{N} \times \mathbf{F_{O}}}$ is the layer output with $\mathbf{F_{O}}$ features, $Y_{j} \in \mathbb{R}^{\mathbf{N}}$ is its $j^{th}$ feature. $W \in \mathbb{R}^{\mathbf{F_{I}} \times \mathbf{F_{O}}}$ is a trainable parameter. $\boldsymbol{\rho}(\boldsymbol{\cdot})$ is the activation function. Such layer can aggregate information of 1-hop neighbors. The receptive field of neighborhood can be expanded by stacking multiple graph convolution layers \cite{chen2019multi}.

\subsubsection{\textbf{Diffusion Graph Convolution}}
Spectral graph convolution requires a symmetric Laplacian matrix to implement eigendecomposition. It becomes invalid for a directed graph with an asymmetric Laplacian matrix. Diffusion convolution origins from graph diffusion and has no constraint on graph. Graph diffusion \cite{DBLP:journals/fttcs/Teng16}, \cite{teng2016scalable} can be represented as a transition matrix power series giving the probability of jumping from one node to another node at each step. After many steps, such Markov process converges to a stationary distribution $\mathcal{P}=\sum_{k=0}^{\infty} \alpha(1-\alpha)^{k}(\mathbf{D_{O}}^{-1}\mathbf{A})^{k}$, where $\mathbf{D_{O}}^{-1}\mathbf{A}$ is the transition matrix, $\alpha \in [0,1]$ is the restart probability and $k$ is the diffusion step. In practice, a finite $\mathbf{K}$-step truncation of the diffusion process is adopted and each step is assigned a trainable weight $\theta$. Based on the $\mathbf{K}$-step diffusion process, Li et al. \cite{DBLP:conf/iclr/LiYS018} defined diffusion graph convolution as:

\begin{footnotesize}\begin{equation}\setlength{\abovedisplayskip}{2pt}\setlength{\belowdisplayskip}{2pt}
y=\Theta \boldsymbol{*_{\mathcal{G}}} x=\sum_{k=0}^{\mathbf{K}-1}(\theta_{k, 1}(\mathbf{D_{O}}^{-1}\mathbf{A})^{k}+\theta_{k, 2}(\mathbf{D_{I}}^{-1} \mathbf{A}^{T})^{k})x
\end{equation}\end{footnotesize}here, $\mathbf{D_{O}}^{-1}\mathbf{A}$ represents the transition matrix and $\mathbf{D_{I}}^{-1} \mathbf{A}^{T}$ is its transpose. Such bidirectional diffusion enables the operation to capture the spatial correlation on a directed graph\cite{DBLP:conf/iclr/LiYS018}. Similar to spectral graph convolution layer, a diffusion graph convolutional layer is built as follows:

\begin{footnotesize}\begin{equation}\setlength{\abovedisplayskip}{2pt}\setlength{\belowdisplayskip}{2pt}
\begin{split}
Y_{j}&=\boldsymbol{\rho}( \sum_{k=0}^{\mathbf{K}-1}\sum_{i=1}^\mathbf{F_{I}}(\theta_{k, 1, i, j}(\mathbf{D_{O}}^{-1}\mathbf{A})^{k}+\theta_{k, 2, i, j}(\mathbf{D_{I}}^{-1} \mathbf{A}^{T})^{k}) X_{i})\\
Y&=\boldsymbol{\rho}(\sum_{k=0}^{\mathbf{K}-1}(\mathbf{D_{O}}^{-1}\mathbf{A})^{k} X W_{k1}+(\mathbf{D_{I}}^{-1} \mathbf{A}^{T})^{k}X W_{k2})  
\end{split}
\end{equation}\end{footnotesize}where $1 \leq j \leq \mathbf{F_{O}}$, parameters $W_{k1},W_{k2} \in \mathbb{R}^{\mathbf{F_{I}} \times \mathbf{F_{O}}}$ are trainable.

\subsubsection{\textbf{GNNs in Traffic Domain}}
Many traffic works, such as subway network and road network,  are graph structure naturally (See Section \ref{sec:GraphConstruction}). Compared with previous works modeling traffic network as grids \cite{zhang2017deep},\cite{DBLP:journals/ijcisys/DuLGH20}, the works modeling traffic network as graph can fully utilize spatial information.

By now, many works employ convolution operation  directly on traffic graph to capture the complex spatial dependency of traffic data. Most of them adopt spectral graph convolution (SGC) while some employ diffusion graph convolution (DGC) \cite{DBLP:conf/aaai/ChenLTZWWZ19}, \cite{DBLP:conf/iclr/LiYS018}, \cite{DBLP:conf/ijcnn/ZhangWCC19}, \cite{DBLP:conf/ijcai/WuPLJZ19}, \cite{DBLP:conf/trustcom/HuangWYC19},\cite{DBLP:journals/corr/abs-1906-00560}. There are also some other graph based deep learning techniques such as graph attention network (GAT) \cite{DBLP:conf/uai/ZhangSXMKY18}, \cite{DBLP:journals/access/ZhangYL19a}, \cite{DBLP:journals/corr/abs-1911-08415},\cite{kanglearning19}, tensor decomposition and completion on graph \cite{DBLP:journals/corr/abs-1912-05693}, but their related works are few, which might be a future research direction.

The key difference between SGC and DGC lies in their matrices which represent different assumptions on the spatial correlations in traffic network. The adjacency matrix in SGC infers that a central node in a graph has stronger correlation with its adjacent nodes than other distant ones \cite{geng2019spatiotemporal},\cite{DBLP:conf/mdm/GeLLZ19}. The state transition matrix in DGC indicates that the spatial dependency is stochastic depending on the restart probability and dynamic instead of being fixed. The traffic flow is related to a diffusion process on a traffic graph to model its dynamic spatial correlations. In addition, the bidirectional diffusion in DGC offers the model more flexibility to capture the influence from both upstream and downstream traffic \cite{DBLP:conf/iclr/LiYS018}. In a word, DGC is more complicated than SGC. DGC can be adopted in both symmetric or asymmetric traffic network graph while SGC can be only utilized to process symmetric traffic graph.

Existing graph convolution theories are mainly applied on 2-D signal $X \in \mathbb{R}^{\mathbf{N} \times \mathbf{F_{I}}}$. However, the traffic data with both spatial and temporal attributes are usually 3-D signal $\mathcal{X}\in \mathbb{R}^{\mathbf{P} \times\mathbf{N} \times \mathbf{F_{I}}}$. The convolution operations need to be further generalized to 3-D signal. Equal convolution operation (e.g. SGC, DGC) with the same kernel is imposed on each time step of 3-D signal $\mathcal{X}$ in parallel \cite{DBLP:conf/ijcai/YuYZ18}, \cite{DBLP:conf/mdm/GeLLZ19}, \cite{DBLP:conf/ijcai/FangZMXP19},\cite{wang2018dynamic}.

In order to enhance the performance of graph convolution in traffic tasks, many works develop various variants of SGC.

Guo et al. \cite{DBLP:conf/aaai/GuoLFSW19} redefined SGC with attention mechanism to adaptively capture the dynamic correlations in traffic network: $\Theta \boldsymbol{*_{\mathcal{G}}} x \approx \sum_{k=0}^{\mathbf{K}-1} \theta_{k}(T_{k}(\tilde{\mathbf{L}})\boldsymbol{\odot} \mathbf{S})x$ , where $\mathbf{S} = W_{1} \boldsymbol{\odot}\boldsymbol{\rho}((XW_{2})W_{3}(W_{4}X)^{T}+b)\in \mathbb{R}^{\mathbf{N} \times \mathbf{N}}$ is the spatial attention.

Yu et al. \cite{DBLP:journals/corr/abs-1903-00919} generalized SGC on both spatial and temporal dimensions by scanning $\mathbf{K}$ order neighbors on graph and $\mathbf{K}_{t}$ neighbors on time-axis without padding. The equation is as follows: 

\begin{footnotesize}\begin{equation}\setlength{\abovedisplayskip}{2pt}\setlength{\belowdisplayskip}{2pt}
\mathcal{Y}_{t,j}=\boldsymbol{\rho}( \sum_{t^{\prime}=0}^{\mathbf{K}_{t}-1} \sum_{k=0}^{\mathbf{K}-1}\sum_{i=1}^{\mathbf{F_{I}}}\theta_{j, t^{\prime}, k, i} \tilde{\mathbf{L}}^{k} \mathcal{X}_{t-t^{\prime},i})
\end{equation}\end{footnotesize}where $\mathcal{X}_{t-t^{\prime},i}\in \mathbb{R}^{\mathbf{N}}$ is the $i^{th}$ feature of input $\mathcal{X}$ at time $t-t^{\prime}$ , $\mathcal{Y}_{t,j}\in \mathbb{R}^{\mathbf{N}}$ is the $j^{th}$ feature of output $\mathcal{Y}$ at time $t$.

Zhao et al. \cite{cuitraffic19} changed SGC as $\Theta \boldsymbol{*_{\mathcal{G}}} x=(W \boldsymbol{\odot} \tilde{\mathbf{A}}^{\mathbf{K}} \boldsymbol{\odot} \mathcal{F F R})x$ , where $\tilde{\mathbf{A}}^{\mathbf{K}}$ is the $\mathbf{K}$-hop neighborhood matrix and $\mathcal{F F R}$ is a matrix representing physical properties of road network. Some researchers \cite{DBLP:journals/corr/abs-1909-07105},\cite{zhang2019multistep} followed this work and redefined $\Theta \boldsymbol{*_{\mathcal{G}}} x=(W \boldsymbol{\odot} Bi(\mathbf{A}^{\mathbf{K}}+\mathbf{I_{N}}))x$, where $Bi(.)$ is a function clipping each nonzero element in matrix to 1.

Sun et al. \cite{DBLP:journals/corr/abs-1903-07789} modified adjacency matrix $\mathbf{A}$ in SGC as $\mathbf{S}= \mathbf{A} \boldsymbol{\odot} \omega$ to integrate the geospatial positions information into the model and $\omega$ is a matrix calculated via a thresholded Gaussian kernel weighting function. The layer is built as $Y=\boldsymbol{\rho}(\tilde{\mathbf{Q}}^{-\frac{1}{2}} \tilde{\mathbf{S}} \tilde{\mathbf{Q}}^{-\frac{1}{2}} X W)$, where $\tilde{\mathbf{Q}}$ is the degree matrix of $\tilde{\mathbf{S}}=\mathbf{S}+\mathbf{I_{N}}$. 

Qiu et al. \cite{qiu2019dynamic} designed a novel edge-based SGC on road network to extract the spatiotemporal correlations of the edge features. Both the feature matrix $X$ and adjacency matrix $\mathbf{A}$ are defined on edges instead of nodes.

\subsection{RNNs}
Recurrent Neural Networks (RNNs) are a type of neural network architecture which is mainly used to detect patterns in sequential data \cite{DBLP:journals/corr/abs-1912-05911}. The traffic data collected in many traffic tasks are time series data, thus  RNNs are commonly utilized in these traffic tasks to capture the temporal dependency in traffic data. In this subsection, we introduce three classical models of RNNs (i.e. RNN, LSTM, GRU) and the correlations among them, providing theoretical evidence for participators to choose appropriate models for specific traffic problems. 

\subsubsection{\textbf{RNN}}
%%%%%%%%%%%%%%%%%%%%% Figures %%%%%%%%%%%%%%%%%%
\begin{figure}[htb]
\centering
\includegraphics[width=0.4\textwidth, height=0.13\textheight]{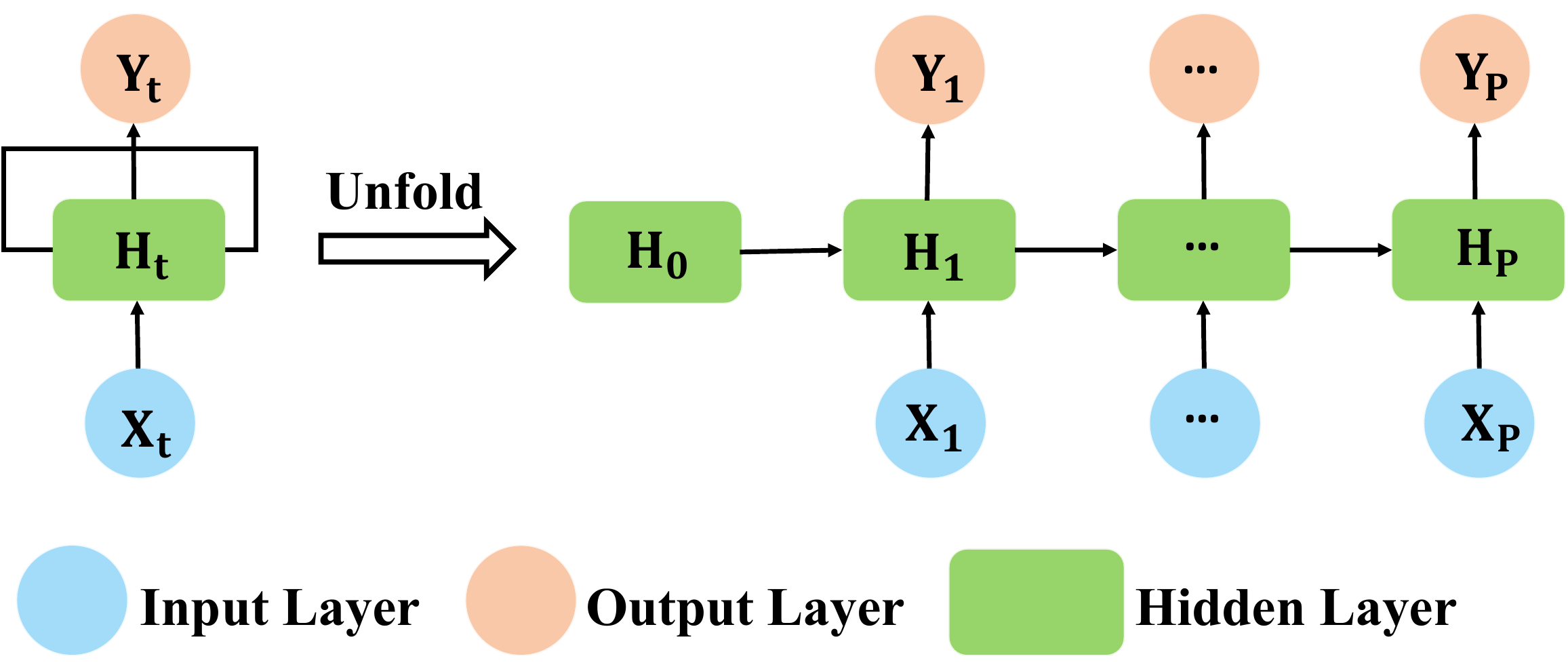}
\caption{The folded and unfolded structure of recurrent neural networks}
\label{fig:RNNs}
\end{figure}

Similar to a classical Feedforward Neural Network (FNN), a simple recurrent neural network (RNN) \cite{DBLP:journals/tnn/BengioSF94} contains three layers, i.e. input layer, hidden layer, output layer \cite{salehinejad2017recent}. What differentiates RNN from FNN is the hidden layer. It passes information forward to the output layer in FNN while in RNN, it also transmits information back into itself forming a cycle \cite{DBLP:journals/corr/abs-1912-05911}. For this reason, the hidden layer in RNN is called recurrent hidden layer. Such cycling trick can retain historical information, enabling RNN to process time series data.

Suppose there are $\mathbf{F_{I}}$, $\mathbf{F_{H}}$, $\mathbf{F_{O}}$ units in the input, hidden, output layer of RNN respectively. The input layer takes time series data $\mathbf{X}=[\mathbf{X}_{1},\cdots,\mathbf{X}_{\mathbf{P}}] \in \mathbb{R}^{\mathbf{P}\times \mathbf{F_{I}}}$ in. For each element $\mathbf{X}_{t}\in \mathbb{R}^{\mathbf{F_{I}}}$ at time $t$, the hidden layer transforms it to $\mathbf{H}_{t}\in \mathbb{R}^{\mathbf{F_{H}}}$ and the output layer maps  $\mathbf{H}_{t}$ to $\mathbf{Y}_{t}\in \mathbb{R}^{\mathbf{F_{O}}}$.
Note that the hidden layer not only takes $\mathbf{X}_{t}$ as input but also takes $\mathbf{H}_{t-1}$ as input. Such cycling mechanism enables RNN to memorize the past information (as shown in Figure \ref{fig:RNNs}). The mathematical notations of hidden layer and output layer are as follows:

\begin{footnotesize}\begin{equation}\setlength{\abovedisplayskip}{2pt}\setlength{\belowdisplayskip}{2pt}
\begin{split}
\mathbf{H}_{t}&=\boldsymbol{tanh}([\mathbf{H}_{t-1},\mathbf{X}_{t}] \boldsymbol{\cdot} W_{h}+b_{h})\\
\mathbf{Y}_{t}&=\boldsymbol{\rho}(\mathbf{H}_{t}\boldsymbol{\cdot} W_{y}+b_{y})
\end{split}
\end{equation}\end{footnotesize}where $W_{h}\in \mathbb{R}^{(\mathbf{F_{I}+\mathbf{F_{H}}})\times \mathbf{F_{H}}}$, $W_{y}\in \mathbb{R}^{\mathbf{F_{H}}\times \mathbf{F_{O}}}$, $b_{h}\in \mathbb{R}^{\mathbf{F_{H}}}$, $b_{y}\in \mathbb{R}^{\mathbf{F_{O}}}$ are trainable parameters. $t=1, \cdots, \mathbf{P}$ and $\mathbf{P}$ is the input sequence length. $\mathbf{H}_{0}$ is initialized using small non-zero elements which can improve overall performance and stability of the network \cite{DBLP:conf/icml/SutskeverMDH13}. 

In a word, RNN takes sequential data as input and generates another sequence with the same length: $[\mathbf{X}_{1},\cdots,\mathbf{X}_{\mathbf{P}}]\stackrel{RNN}{\longrightarrow}[\mathbf{Y}_{1},\cdots,\mathbf{Y}_{\mathbf{P}}]$. Note that we can deepen RNN through stacking multiple recurrent hidden layers.

\subsubsection{\textbf{LSTM}}
Although the hidden state enables RNN to memorize the input information over past time steps, it also introduces matrix multiplication over the (potentially very long) sequence. Small values in the matrix multiplication cause the gradients to decrease at each time step, resulting in final vanish phenomenon. Oppositely big values lead to exploding problem \cite{chen2016gentle}. The vanishing or exploding gradients actually hinder the capacity of RNN to learn long-term sequential dependencies in data \cite{salehinejad2017recent}.

To overcome this hurdle, Long Short-Term Memory (LSTM) neural networks\cite{hochreiter1997long} are proposed to capture long-term dependency in sequence learning. Compared with the hidden layer in RNN, LSTM hidden layer has extra four parts,  i.e. a memory cell, input gate, forget gate, and output gate. These three gates ranging in [0,1] can control information flow into the memory cell and preserve the extracted features from previous time steps. These simple changes enable the memory cell to store and read as much long-term information as possible. The mathematical notations of LSTM hidden layer are as follows:

\begin{footnotesize}\begin{equation}\setlength{\abovedisplayskip}{2pt}\setlength{\belowdisplayskip}{2pt}
\begin{split}
i_{t}&=\boldsymbol{\sigma}([\mathbf{H}_{t-1},\mathbf{X}_{t}]\boldsymbol{\cdot} W_{i}+b_{i})\\
o_{t}&=\boldsymbol{\sigma}([\mathbf{H}_{t-1},\mathbf{X}_{t}]\boldsymbol{\cdot} W_{o}+b_{o})\\
f_{t}&=\boldsymbol{\sigma}([\mathbf{H}_{t-1},\mathbf{X}_{t}]\boldsymbol{\cdot} W_{f}+b_{f})\\
\mathbf{C}_{t}&=f_{t} \boldsymbol{\odot} \mathbf{C}_{t-1}+i_{t} \boldsymbol{\odot} \boldsymbol{tanh}([\mathbf{H}_{t-1},\mathbf{X}_{t}]\boldsymbol{\cdot} W_{c}+b_{c})\\
\mathbf{H}_{t}&=o_{t} \boldsymbol{\odot} \boldsymbol{tanh}(\mathbf{C}_{t})
\end{split}
\end{equation}\end{footnotesize}where $i_{t}$, $o_{t}$, $f_{t}$ are the input gate, output gate, forget gate at time $t$ respectively. $\mathbf{C}_{t}$ is the memory cell at time $t$.  

\subsubsection{\textbf{GRU}}
While LSTM is a viable option for avoiding vanishing or exploding gradients, its complex structure leads to more memory requirement and longer training time. Chung et al. \cite{chung2014empirical} proposed a simple yet powerful variant of LSTM, i.e. Gated Recurrent Unit (GRU). The LSTM cell has three gates, but the GRU cell only has two gates, resulting in fewer parameters thus shorter training time. However, GRU is equally effective as LSTM empirically \cite{chung2014empirical} and is widely used in various tasks. The mathematical notations of GRU hidden layer are as follows:

\begin{footnotesize}\begin{equation}\setlength{\abovedisplayskip}{2pt}\setlength{\belowdisplayskip}{2pt}
\begin{split}
r_{t}&=\boldsymbol{\sigma}([\mathbf{H}_{t-1},\mathbf{X}_{t}]\boldsymbol{\cdot} W_{r}+b_{r})\\
u_{t}&=\boldsymbol{\sigma}([\mathbf{H}_{t-1},\mathbf{X}_{t}]\boldsymbol{\cdot} W_{u}+b_{u})\\
\tilde{\mathbf{H}_{t}}&=
\boldsymbol{tanh}(r_{t} \boldsymbol{\odot} [\mathbf{H}_{t-1},\mathbf{X}_{t}]\boldsymbol{\cdot} W_{h}+b_{h})\\
\mathbf{H}_{t}&= u_{t} \boldsymbol{\odot} \mathbf{H}_{t-1} +(1-u_{t})\boldsymbol{\odot} \tilde{\mathbf{H}_{t}} \\
\end{split}
\end{equation}\end{footnotesize}where $r_{t}$ is the reset gate,  $u_{t}$ is the update gate. 

\subsubsection{\textbf{RNNs in Traffic Domain}}
RNNs have shown impressive capability of processing time series data. Since traffic data has a distinct temporal dependency, RNNs are usually leveraged to capture temporal correlation in traffic data. Among the works we survey, only Geng et al. \cite{geng2019spatiotemporal} utilized RNN to capture temporal dependency in traffic data while more than a half adopted GRU and some employed LSTM. This can be explained that RNN survives severe gradient disappearance or gradient explosion while LSTM and GRU handle this successfully and GRU can reduce the training time. 

In addition, there are many tricks to augment RNNs' capacity to model the complex temporal dynamics in traffic domain, such as attention mechanism, gating mechanism and residual mechanism.

For instance, Geng et al. \cite{geng2019spatiotemporal} incorporated the contextual information, i.e. output of SGCN which contains information of related regions, into an attention operation to model the correlations between observations at different timestamps: 

\begin{footnotesize}\begin{equation}\setlength{\abovedisplayskip}{2pt}\setlength{\belowdisplayskip}{2pt}
\begin{split}
z &= F_{pool}(\mathbf{X}_{t},SGCN(\mathbf{X}_{t}))\\
S &= \boldsymbol{\sigma} (W_{1}\boldsymbol{ReLU}(W_{2}z))\\
\mathbf{H}_{t} &= RNN([\mathbf{H}_{t-1},\mathbf{X}_{t}]\boldsymbol{\odot} S)\\
\end{split}
\end{equation}\end{footnotesize}where $F_{pool}(\boldsymbol{\cdot})$ is a global average pooling layer, $RNN(\boldsymbol{\cdot})$ denotes the RNN hidden layer.

Chen et al. \cite{DBLP:conf/aaai/ChenLTZWWZ19} took external factors into consideration by embedding external attributes into the input. In addition, they added the previous hidden states to the next hidden states through a residual shortcut path, which they believed can make GRU more sensitive and robust to sudden changes in traffic historical observations. The new hidden state is formulated as: $\mathbf{H}_{t} = GRU([\mathbf{H}_{t-1},\mathbf{X}_{t}],\mathbf{E}_{t})+\mathbf{H}_{t-1}W$, where $\mathbf{E}_{t}$ is the external features at time $t$, $W$ is a linear trainable parameter, $\mathbf{H}_{t-1}W$ is the residual shortcut.

Yu et al. \cite{DBLP:journals/corr/abs-1903-05631} inserted a dilated skip connection into GRU by changing hidden state from $\mathbf{H}_{t} = GRU([\mathbf{H}_{t-1},\mathbf{X}_{t}])$ to $\mathbf{H}_{t} = GRU(\mathbf{H}_{t-s},\mathbf{X}_{t})$, where $s$ refers to the skip length or dilation rate of each layer, $GRU(\boldsymbol{\cdot})$ denotes the GRU hidden layer. Such hierarchical design of dilation brings in multiple temporal scales for recurrent units at different layers which achieves multi-timescale modeling.

Despite the tricks above, some works replace the matrix multiplication in RNNs' hidden layer with spectral graph convolution (SGC) or diffusion graph convolution (DGC), to capture spatial-temporal correlations jointly. Take GRU as example:

\begin{footnotesize}\begin{equation}\setlength{\abovedisplayskip}{2pt}\setlength{\belowdisplayskip}{2pt}
\begin{split}
r_{t}&=\boldsymbol{\sigma}([\mathbf{H}_{t-1},\mathbf{X}_{t}] \boldsymbol{*_{\mathcal{G}}} W_{r}+b_{r})\\
u_{t}&=\boldsymbol{\sigma}([\mathbf{H}_{t-1},\mathbf{X}_{t}] \boldsymbol{*_{\mathcal{G}}} W_{u}+b_{u})\\
\tilde{\mathbf{H}_{t}}&=\boldsymbol{tanh}(r_{t} \boldsymbol{\odot} [\mathbf{H}_{t-1},\mathbf{X}_{t}]\boldsymbol{*_{\mathcal{G}}} W_{h}+b_{h})\\
\mathbf{H}_{t}&= u_{t} \boldsymbol{\odot} \mathbf{H}_{t-1} + (1-u_{t}) \boldsymbol{\odot}\tilde{\mathbf{H}_{t}} \\
\end{split}
\end{equation}\end{footnotesize}The $\boldsymbol{*_{\mathcal{G}}}$ can represent SGC, DGC or other convolution operations. In the literatures we survey, most replacements happen in GRU and only one in LSTM \cite{DBLP:conf/kdd/LiHCSWZP19}. Among GRU related traffic works, \cite{DBLP:conf/aaai/ChenLTZWWZ19}, \cite{DBLP:conf/iclr/LiYS018}, \cite{guooptimized20}, \cite{DBLP:conf/trustcom/HuangWYC19},\cite{DBLP:journals/corr/abs-1906-00560} replaced matrix multiplication with DGC, \cite{chen2019multi}, \cite{DBLP:journals/corr/abs-1903-05631}, \cite{DBLP:journals/corr/abs-2001-04889} with SGC, \cite{kanglearning19}, \cite{DBLP:conf/uai/ZhangSXMKY18} with GAT.

Note that besides RNNs, other techniques (e.g. TCN in the next subsection) are also popular choices to extract the temporal dynamics in traffic tasks.

\subsection{TCN}
Although RNN-based models become widespread in time-series analysis, RNNs for traffic prediction still suffer from time-consuming iteration, complex gate mechanism, and slow response to dynamic changes \cite{DBLP:conf/ijcai/YuYZ18}. On the contrary, 1D-CNN has the superiority of fast training, simple structure, and no  constraints to previous steps \cite{kalchbrenner2016neural}. However, 1D-CNN is less common than RNNs in practice due to its lack of memory for a long sequence \cite{DBLP:conf/icml/DauphinFAG17}. In 2016, a novel convolution operation integrating causal convolution and dilated convolution \cite{DBLP:conf/ssw/OordDZSVGKSK16} is proposed,  which outperforms RNNs in text-to-speech tasks. The prediction of causal convolution depends on previous elements but not on future elements. Dilated convolution expands the receptive field of original filter by dilating it with zeros \cite{DBLP:journals/corr/YuK15}. Bai et al. \cite{DBLP:journals/corr/abs-1803-01271} simplified the causal dilated convolution \cite{DBLP:conf/ssw/OordDZSVGKSK16} for sequence modeling problem and renamed it as temporal convolution network (TCN). Recently, more and more works employ TCN to process traffic data \cite{DBLP:conf/ijcai/YuYZ18}, \cite{DBLP:conf/mdm/GeLLZ19}, \cite{DBLP:conf/ijcai/WuPLJZ19}, \cite{DBLP:conf/ijcai/FangZMXP19}.

\subsubsection{\textbf{Sequence Modeling and 1-D TCN}}
Given an input sequence with length $\mathbf{P}$ denoted as $\mathbf{x}=[\mathbf{x}_{1},\cdots,\mathbf{x}_{\mathbf{P}}] \in \mathbb{R}^{\mathbf{P}}$, sequence modeling aims to generate an output sequence with the same length, denoted as $\mathbf{y}=[\mathbf{y}_{1},\cdots,\mathbf{y}_{\mathbf{P}}]\in \mathbb{R}^{\mathbf{P}}$. The key assumption is that the output at current time $\mathbf{y}_{t}$ only depends on historical data $[\mathbf{x}_{1},\cdots,\mathbf{x}_{t}]$ but does not depend on any future inputs $[\mathbf{x}_{t+1},\cdots,\mathbf{x}_{\mathbf{P}}]$, i.e. $\mathbf{y}_{t}=f(\mathbf{x}_{1},\cdots,\mathbf{x}_{t})$, $f$ is the mapping function.

\begin{figure}[htb]
\centering
\includegraphics[width=0.45\textwidth, height=0.15\textheight]{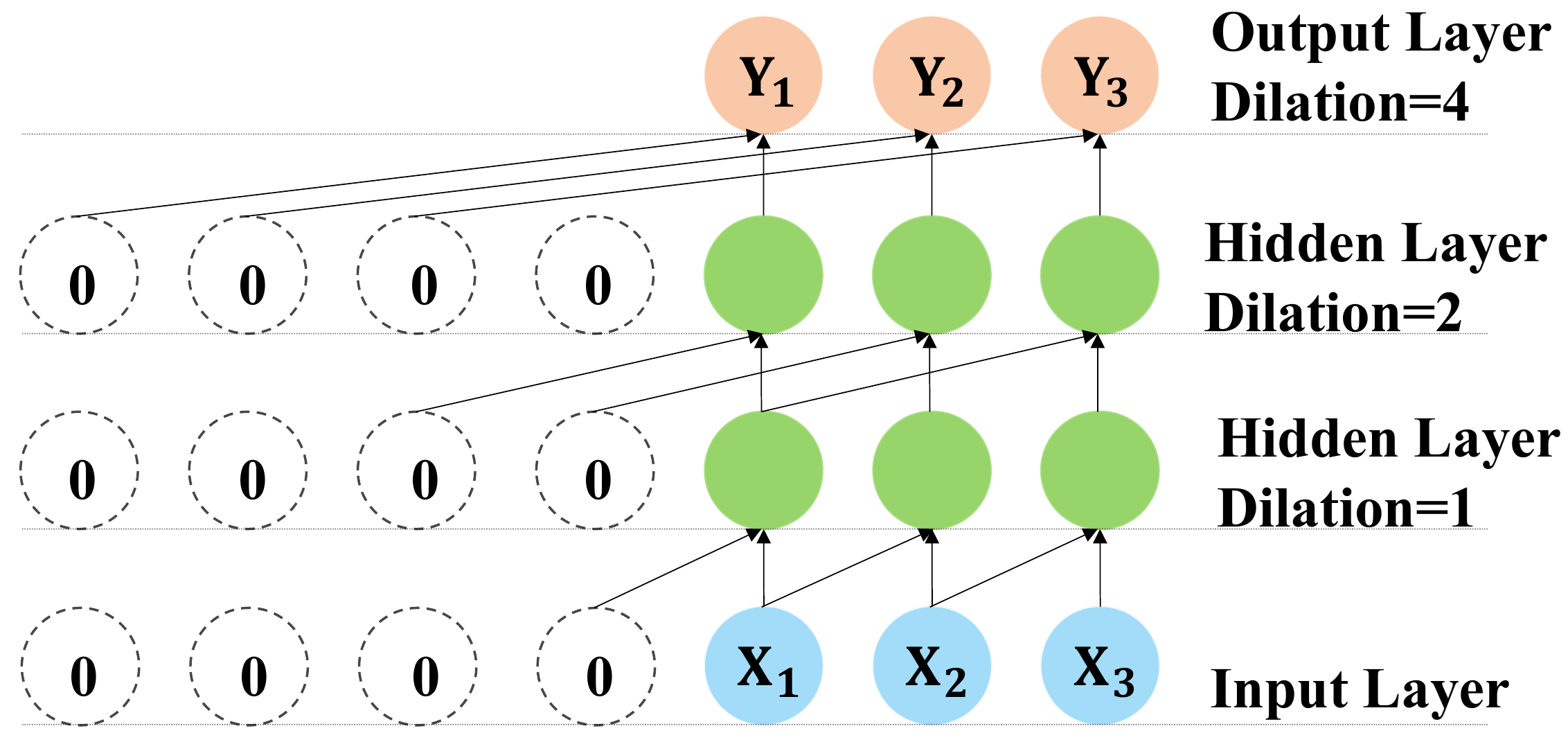}
\caption{Multiple dilated causal convolution layers in TCN: $[\mathbf{x}_{1},\mathbf{x}_{2},\mathbf{x}_{3}]$ is the input sequence and $[\mathbf{y}_{1},\mathbf{y}_{2},\mathbf{y}_{3}]$ is the output sequence with the same length. The size of kernel is $2$ and the dilation rate sequence is $[1,2,4]$. Zero padding strategy is taken.}
\label{fig:TCN}
\end{figure}

Obviously, RNN, LSTM and GRU can be solutions to sequence modeling tasks. However, TCN can tackle sequence modeling problem more efficiently than RNNs for that it can capture long sequence properly in a non-recursive manner. The dilated causal convolution in TCN is formulated as follows:

\begin{footnotesize}\begin{equation}\setlength{\abovedisplayskip}{2pt}\setlength{\belowdisplayskip}{2pt}
\mathbf{y}_{t}=\Theta *_{\mathcal{T}^{\mathbf{d}}} \mathbf{x}_{t} =\sum_{k=0}^{\mathbf{K}-1} w_{k} \mathbf{x}_{t-\mathbf{d}k}
\end{equation}\end{footnotesize}where $*_{\mathcal{T}^{\mathbf{d}}}$ is the dilated causal operator with dilation rate $\mathbf{d}$ controlling the skipping distance, $\Theta=[w_{0}, \cdots, w_{\mathbf{K-1}}]\in \mathbb{R}^{\mathbf{K}}$ is the kernel. Zero padding strategy is utilized to keep the output length the same as the input length (as shown in Figure \ref{fig:TCN}). Without  padding, the output length is shortened by $(\mathbf{K}-1)\mathbf{d}$ \cite{DBLP:conf/ijcai/YuYZ18}.
%%%%%%%%%%%%%%%%%%%%% Figures %%%%%%%%%%%%%%%%%%

To enlarge the receptive field, TCN stacks multiple dilated causal convolution layers with $\mathbf{d}=2^{l}$ as the dilation rate of $l^{th}$ layer (as shown in Figure  \ref{fig:TCN}). Therefore, the receptive field in the network grows exponentially without requiring many convolutional layers or larger filter, which can handle longer sequence with less layers and save computation resources \cite{DBLP:conf/ijcai/WuPLJZ19}.

\subsubsection{\textbf{TCN in Traffic Domain}}
There are many traffic works related with sequence modeling, especially traffic spatial-temporal forecasting tasks. Compared with RNNs, the non-recursive calculation manner enables TCN to alleviate the gradient explosion problem and facilitate the training by parallel computation. Therefore, some works adopt TCN to capture the temporal dependency in traffic data.

Most graph-based traffic data is 3-D signal denoted as $\mathcal{X}\in \mathbb{R}^{\mathbf{P}\times \mathbf{N} \times \mathbf{F_{I}}}$, which requires the generalization of 1-D TCN to 3-D TCN. The dilated causal convolution can be adopted to produce the $j^{th}$ output feature of node $i$ at time $t$ as follows \cite{DBLP:conf/mdm/GeLLZ19}:

\begin{footnotesize}\begin{equation}\setlength{\abovedisplayskip}{2pt}\setlength{\belowdisplayskip}{2pt}
\begin{split}
\mathcal{Y}_{t,j}^{i}&=\boldsymbol{\rho}(\Theta_{j}*_{\mathcal{T}^{\mathbf{d}}}\mathcal{X}_{t}^{i})
=\boldsymbol{\rho}(\sum_{m=1}^{\mathbf{F_{I}}}\sum_{k=0}^{\mathbf{K}-1}w_{j,m,k}\mathcal{X}_{t-\mathbf{d}k,m}^{i})
\end{split}
\end{equation}\end{footnotesize}where $1 \leq j \leq \mathbf{F_{O}}$, $\mathcal{Y}_{t,j}^{i}\in \mathbb{R}$ is the $j^{th}$ output feature of node $i$ at time $t$. $\mathcal{X}_{t-\mathbf{d}k,m}^{i}\in \mathbb{R}$ is the $m^{th}$ input feature of node $i$ at time $t-\mathbf{d}k$. The kernel $\Theta_{j}\in \mathbb{R}^{\mathbf{K} \times \mathbf{F_{I}}}$ is trainable. $\mathbf{F_{O}}$ is the number of output features.

The same convolution kernel is applied to all nodes in the traffic network and each node produces $\mathbf{F_{O}}$ new features. The mathematical formulation of each layer is as follows \cite{DBLP:conf/mdm/GeLLZ19},\cite{DBLP:conf/ijcai/FangZMXP19}:

\begin{footnotesize}\begin{equation}\setlength{\abovedisplayskip}{2pt}\setlength{\belowdisplayskip}{2pt}
\mathcal{Y}=\boldsymbol{\rho}(\Theta *_{\mathcal{T}^{\mathbf{d}}} \mathcal{X})
\end{equation}\end{footnotesize}where $\mathcal{X}\in \mathbb{R}^{\mathbf{P}\times \mathbf{N} \times \mathbf{F_{I}}}$ represents the historical observations of the whole traffic network over past $\mathbf{P}$ time slices,  $\Theta\in \mathbb{R}^{\mathbf{K} \times \mathbf{F_{I}}\times \mathbf{F_{O}}}$ represents the related convolution kernel, $\mathcal{Y}\in \mathbb{R}^{\mathbf{P}\times \mathbf{N} \times \mathbf{F_{O}}}$ is the output of TCN layer.

There are some tricks to enhance the performance of TCN in specific traffic tasks. For instance, Fang et al. \cite{DBLP:conf/ijcai/FangZMXP19} stacked multiple TCN layers to extract the short-term neighboring dependency by bottom layer and long-term temporal dependency by higher layer: 

\begin{footnotesize}\begin{equation}\setlength{\abovedisplayskip}{2pt}\setlength{\belowdisplayskip}{2pt}
\mathcal{Y}^{(l+1)}=\boldsymbol{\sigma}(\Theta^{l} *_{\mathcal{T}^{\mathbf{d}}} \mathcal{Y}^{(l)})
\end{equation}\end{footnotesize}where $\mathcal{Y}^{(l)}$ is the input of $l^{th}$ layer, $\mathcal{Y}^{(l+1)}$ is the output and $\mathcal{Y}^{(0)}=\mathcal{X}$. $\mathbf{d}=2^{l}$ is the dilation rate of  $l^{th}$ layer.

To reduce the complexity of model training, Ge et al. \cite{DBLP:conf/mdm/GeLLZ19} constructed a residual block containing two TCN layers with the same dilation rate. The block input was added to last TCN layer to get the block output:

\begin{footnotesize}\begin{equation}\setlength{\abovedisplayskip}{2pt}\setlength{\belowdisplayskip}{2pt}
\mathcal{Y}^{(l+1)}=\mathcal{Y}^{(l)}+\boldsymbol{ReLU}(\Theta_{1}^{l} *_{\mathcal{T}^{\mathbf{d}}} (\boldsymbol{ReLU}(\Theta_{0}^{l} *_{\mathcal{T}^{\mathbf{d}}} \mathcal{Y}^{(l)})))
\end{equation}\end{footnotesize}where $\Theta^{l}_{1}, \Theta^{l}_{2}$ are the convolution kernels of the first layer and the second layer respectively. $\mathcal{Y}^{(l)}$ is the input of residual block and $\mathcal{Y}^{(l+1)}$ is its output.

Wu et al. \cite{DBLP:conf/ijcai/WuPLJZ19} integrated gating mechanism\cite{DBLP:conf/icml/DauphinFAG17} with TCN to learn complex temporal dependency in traffic data:

\begin{footnotesize}\begin{equation}\setlength{\abovedisplayskip}{2pt}\setlength{\belowdisplayskip}{2pt}
\mathcal{Y}=\boldsymbol{\rho}_{1}(\Theta_{1} *_{\mathcal{T}^{\mathbf{d}}} \mathcal{X}+b_{1})\boldsymbol{\odot}\boldsymbol{\rho}_{2}(\Theta_{2} *_{\mathcal{T}^{\mathbf{d}}} \mathcal{X}+b_{2})
\end{equation}\end{footnotesize}where $\boldsymbol{\rho}_{2}(\boldsymbol{\cdot})\in [0,1]$ determines the ratio of information passed to the next layer. 

Similarly, Yu et al. \cite{DBLP:conf/ijcai/YuYZ18} used the Gated TCN and set the dilation rate $\mathbf{d}=1$ without zero padding to shorten the output length as $\mathcal{Y}=(\Theta_{1} *_{\mathcal{T}^{1}} \mathcal{X})\boldsymbol{\odot}\boldsymbol{\sigma}(\Theta_{2} *_{\mathcal{T}^{1}} \mathcal{X})$. They argued that this can discover variances in time series traffic data.

\subsection{Seq2Seq}
\subsubsection{\textbf{Seq2Seq}}
%%%%%%%%%%%%%%%%%%%%% Figures %%%%%%%%%%%%%%%%%%
\begin{figure}[htb]
\centering
\includegraphics[width=0.45\textwidth, height=0.13\textheight]{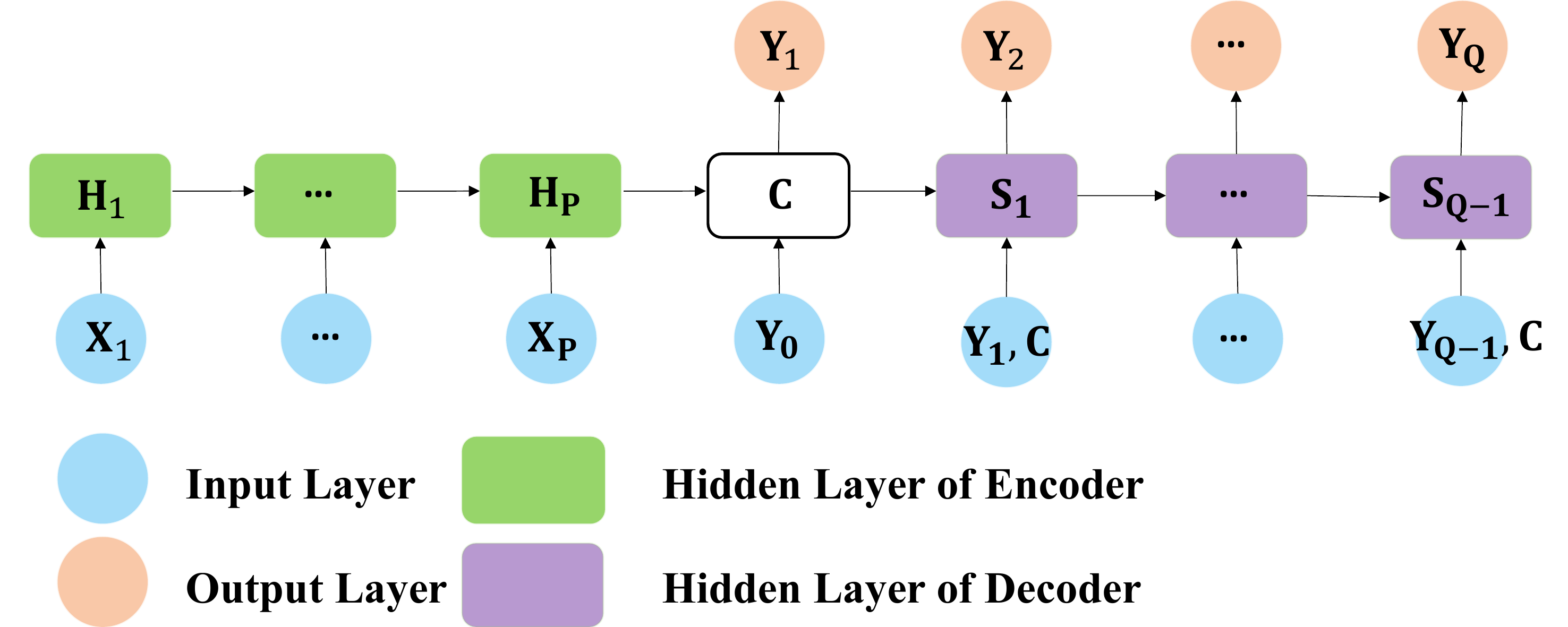}
\caption{Sequence to Sequence Structure without attention mechanism}
\label{fig:Seq2Seq}
\end{figure}
Sequence to Sequence (Seq2Seq) model proposed in 2014 \cite{DBLP:conf/nips/SutskeverVL14} has been widely used in sequence prediction such as machine translation \cite{DBLP:journals/corr/BahdanauCB14}. Seq2Seq architecture consists of two components, i.e. an encoder in charge of converting the input sequence $\mathbf{X}$ into a fixed latent vector $\mathbf{C}$, and a decoder responsible for converting $\mathbf{C}$ into an output sequence $\mathbf{Y}$  (as shown in Figure \ref{fig:Seq2Seq}). Note that $\mathbf{X}$ and $\mathbf{Y}$ can have different lengths.

\begin{footnotesize}\begin{equation}\setlength{\abovedisplayskip}{2pt}\setlength{\belowdisplayskip}{2pt}
\mathbf{X}\!=\![\mathbf{X}_{1}\!,\cdots,\!\mathbf{X}_{i},\cdots,\mathbf{X}_\mathbf{P}]\stackrel{Seq2Seq}{\longrightarrow}\mathbf{Y}=[\mathbf{Y}_{1},\cdots,\mathbf{Y}_{j},\cdots,\mathbf{Y}_\mathbf{Q}]
\end{equation}\end{footnotesize}where $\mathbf{P}$ is the input length and $\mathbf{Q}$ is the output length. $\mathbf{X}_{i} $ is the input at time step $i$. $\mathbf{Y}_{j} $ is the output at time step $j$.

The specific calculation of $\mathbf{Y}_{j}$ is denoted as follows:

\begin{footnotesize}\begin{equation}\setlength{\abovedisplayskip}{2pt}\setlength{\belowdisplayskip}{2pt}
\begin{split}
\mathbf{H}_{i}&=Encoder(\mathbf{X}_{i},\mathbf{H}_{i-1})\\
\mathbf{C}&=\mathbf{H}_{\mathbf{P}},\mathbf{S}_{0}=\mathbf{H}_{\mathbf{P}}\\
\mathbf{S}_{j}&=Decoder(\mathbf{C},\mathbf{Y}_{j-1},\mathbf{S}_{j-1})\\
\mathbf{Y}_{j}&=\mathbf{S}_{j}W
\end{split}
\end{equation}\end{footnotesize}here, $\mathbf{H}_{i}$ is the hidden state of encoder. $\mathbf{H}_{0}$ is initialized using small non-zero elements. $\mathbf{S}_{j}$ is the decoder hidden state. $\mathbf{Y}_{0}$ is the representation of beginning sign. Note that the encoder and decoder can be any model as long as it can accept sequence and produce sequence, such as RNN, LSTM, GRU or other novel models.

A major limitation of Seq2Seq is that the latent vector $\mathbf{C}$ is fixed for each $\mathbf{Y}_{j}$ while $\mathbf{Y}_{j}$ might have stronger correlation with $\mathbf{X}_{j}$ than other elements. To address this issue, attention mechanism is integrated into Seq2Seq, allowing the decoder to focus on task-relevant parts of the input sequence, helping the decoder make better prediction.

\begin{footnotesize}\begin{equation}\setlength{\abovedisplayskip}{2pt}\setlength{\belowdisplayskip}{2pt}
\begin{split}
\mathbf{H}_{i}&=Encoder(\mathbf{X}_{i},\mathbf{H}_{i-1})\\
\mathbf{C}_{j}&=\sum_{i=1}^{\mathbf{P}}(\theta_{ji}\mathbf{H}_{i}),\mathbf{S}_{0}=\mathbf{H}_{\mathbf{P}}\\
\mathbf{S}_{j}&=Decoder(\mathbf{C}_{j},\mathbf{Y}_{j-1},\mathbf{S}_{j-1})\\
\mathbf{Y}_{j}&=\mathbf{S}_{j}W\\
\end{split} 
\end{equation}\end{footnotesize}where $\theta_{ji}=\frac{\exp (f_{ji})}{\sum_{k=1}^{\mathbf{P}} \exp(f_{jk})}$ is the normalized attention score, and $f_{ji}=f(\mathbf{H}_{j},\mathbf{S}_{i-1})$ \cite{DBLP:journals/corr/BahdanauCB14} is a function to measure the correlation between $i^{th}$ input and $j^{th}$ output, for instance, Luong et al. \cite{DBLP:conf/emnlp/LuongPM15} proposed three kinds of attention score calculation.

\begin{footnotesize}\begin{equation}\setlength{\abovedisplayskip}{2pt}\setlength{\belowdisplayskip}{2pt}
f_{ji}=\left\{\begin{array}{ll}\mathbf{H}_{j}^{T} \mathbf{S}_{i-1} & \text { dot } \\ \mathbf{H}_{j}^{T} \boldsymbol{W}_{\boldsymbol{a}} \mathbf{S}_{i-1} & \text { general } \\ \boldsymbol{v}_{a}^{T} \tanh \left(\boldsymbol{W}_{\boldsymbol{a}}\left[\mathbf{H}_{j}, \mathbf{S}_{i-1}\right]\right) & \text { concat }\end{array}\right.   
\end{equation}\end{footnotesize}

Another way to enhance Seq2Seq performance is the scheduled sampling technique \cite{DBLP:conf/nips/BengioVJS15}. The inputs of decoder during training and testing phases are different. Decoder during training phase is fed with true labels of training datasets while it is fed with predictions generated by itself during testing phase, which accumulates error at testing time and causes degraded performance. To mitigate this issue, scheduled sampling is integrated into the model. At $j^{th}$ iteration during the training process,  the probability of feeding the decoder with true label is set as $\epsilon_{j}$ and the probability of feeding the decoder with prediction at the previous step is set as $1-\epsilon_{j}$. Probability $\epsilon_{j}$ gradually decreases to 0, allowing the decoder to learn the testing distribution \cite{DBLP:conf/iclr/LiYS018}, keeping the training and testing as same as possible.

\subsubsection{\textbf{Seq2Seq in Traffic Domain}}
Since Seq2Seq can take in an input sequence and generate an output sequence with different length, it is applied on multi-step prediction in many traffic tasks. The encoder encodes the historical traffic data into a latent space vector. Then, the latent vector is fed into a decoder to generate the future traffic conditions. 

Attention mechanism is usually incorporated into Seq2Seq to model the different influence on future prediction from previous traffic observations at different time slots \cite{DBLP:conf/ijcnn/ZhangWCC19},\cite{DBLP:journals/corr/abs-1911-08415}, \cite{zhang2019multistep},\cite{DBLP:conf/ijcai/BaiYK0S19}. 

The encoder and decoder in many traffic literatures are in charge of capturing spatial-temporal dependencies. For instance, Li et al. \cite{DBLP:conf/iclr/LiYS018} proposed DCGRU to be the encoder and decoder, which can capture spatial and temporal dynamics jointly. The design of encoder and decoder is usually the core contribution and novel part of relative works. Note that the encoder and decoder are not necessarily the same and we have made a summarization of  Seq2Seq structure in previous graph-based traffic works (as shown in Table \ref{tab:seq2seq}).
%%%%%%%%%%%%%%%%%%%%%%%%% Table %%%%%%%%%%%%%%%%%%
\begin{table}[htb]
\caption{The encoders and decoders of sequence to sequence architecture}
\centering
\scriptsize
\begin{tabular}{c|c|c}
\toprule
\textbf{References} &\textbf{Encoder}  &\textbf{Decoder}\\ 
\hline
\cite{DBLP:conf/iclr/LiYS018}&GRU+DGCN&Same as encoder\\\hline
\cite{DBLP:conf/ijcnn/ZhangWCC19}&SGCN +LSTM&LSTM+SGCN\\\hline
\cite{DBLP:journals/corr/abs-1911-08415}&STAtt Block&Same as encoder\\\hline
\cite{weidual19}&MLPs&An MLP\\\hline
\cite{DBLP:journals/corr/abs-1903-06261}&SGCN+Pooling+GRU&GCN+Upooling+GRU\\\hline
\cite{kanglearning19}&GRU with graph self-attention&Same as encoder\\\hline
\cite{chen2019multi}&GRU+SGCN&Same as encoder\\\hline
\cite{zhang2019multistep}&SGCN+ bidirectional GRU&Same as encoder\\\hline
\cite{DBLP:conf/ijcai/BaiYK0S19}&Long-term encoder (Gated SGCN)&Short-term encoder\\\hline
\cite{DBLP:conf/cikm/BaiYK0LY19}&SGCN+LSTM&LSTM\\\hline
\cite{DBLP:conf/trustcom/HuangWYC19}&SGCN+GRU&Same as encoder\\\hline
\cite{DBLP:journals/corr/abs-2001-04889}&CGRM (GRU, SGCN)&Same as encoder\\\hline
\cite{DBLP:journals/corr/abs-1910-09103}&LSTM+RGC&RGC\\\hline
\cite{DBLP:conf/gis/ChaiWY18}&LSTM&Same as encoder\\
\bottomrule
\end{tabular}
\label{tab:seq2seq}
\end{table}

RNNs-based decoder has a severe error accumulation problem during testing inference due to that each previous predicted step is the input to produce the next step prediction. The scheduled sampling to alleviate this problem is adopted in 
\cite{DBLP:conf/iclr/LiYS018},\cite{kanglearning19}. RNNs-based decoder is replaced with a short-term and long-term decoder to take in last step prediction exclusively, thus easing error accumulation \cite{DBLP:conf/ijcai/BaiYK0S19}. The utilization of Seq2Seq technique in traffic domain is flexible. For instance,  Seq2Seq is integrated into a bigger framework, being the generator and discriminator of GAN \cite{DBLP:conf/ijcnn/ZhangWCC19}.

\subsection{GAN}

\subsubsection{\textbf{GAN}}
%%%%%%%%%%%%%%%%%%%%% Figures %%%%%%%%%%%%%%%%%%
\begin{figure}[htb]
\centering
\includegraphics[width=0.45\textwidth, height=0.15\textheight]{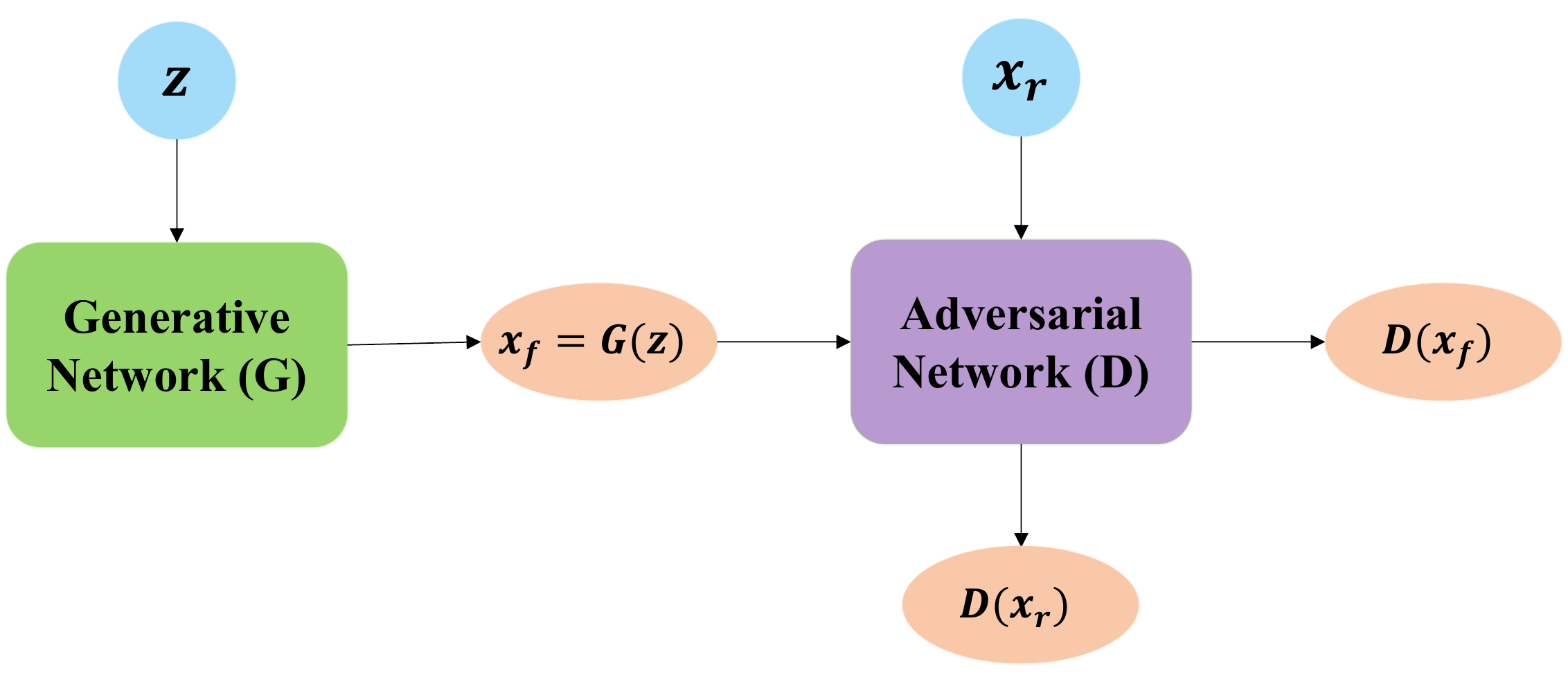}
\caption{Generative Adversarial Network: Generator $G$ is in charge of producing a generated sample $x_{f}=G(z)$ from a random vector $z$, which is sampled from a prior distribution $p_{z}$. Discriminator $D$ is in charge of discriminating between the fake sample $x_{f}$ generated from $G$ and the real sample $x_{r}$ from the training data. }
\label{fig:GAN}
\end{figure}
Generative Adversarial Network (GAN) \cite{goodfellow2014generative} is a powerful deep generative model aiming to generate artificial samples as indistinguishable as possible from their real counterparts. GAN, inspired by game theory, is composed of two players, a generative neural network called Generator $G$ and an adversarial network called Discriminator $D$ (as shown in Figure \ref{fig:GAN}).  

Discriminator $D$ tries to determine whether the input samples belong to the generated data or the real data while Generator $G$ tries to cheat on Discriminator $D$ by producing samples as true as possible. The two mutually adversarial and optimized processes are alternately trained, which strengthens the performance of both $D$ and $G$. When the fake sample produced by $G$ is very close to the ground truth and  $D$ is unable to distinguish them any more, it is considered that Generator $G$ has learned the true distribution of the real data and the model converges. At this time, we can consider this game to reach a Nash equilibrium \cite{DBLP:conf/nips/HeuselRUNH17}.

Mathematically, such process can be formulated to minimize their losses $Loss_{G}$ and $Loss_{D}$. With the loss function being cross entropy denoted as $f$, we can have:

\begin{footnotesize}\begin{equation}\setlength{\abovedisplayskip}{2pt}\setlength{\belowdisplayskip}{2pt}
\begin{split}
Loss_{G}&=f(D(G(z)),1)=-\sum \log D(G(z))\\
\phi^{*}&=\underset{\phi}{\operatorname{argmin}} (Loss_{G})=\underset{\phi}{\operatorname{argmax}} (-Loss_{G})\\
&=\underset{\phi}{\operatorname{argmax}}\mathbb{E}(\log D(G(z)))
\end{split}
\end{equation}\end{footnotesize}

\begin{footnotesize}\begin{equation}\setlength{\abovedisplayskip}{2pt}\setlength{\belowdisplayskip}{2pt}
\begin{split}
Loss_{D}&=f(D(x_{r}), 1, D(x_{f}), 0)\\
        &=-\sum \log D(x_{r})-\sum \log (1-D(x_{f}))\\
\theta^{*}&=\underset{\theta}{\operatorname{argmin}}(Loss_{D})=\underset{\theta}{\operatorname{argmax}} (-Loss_{D})\\
&=\underset{\theta}{\operatorname{argmax}}(\mathbb{E}(\log D(x_{r})+\log(1-D(x_{f}))))
\end{split}
\end{equation}\end{footnotesize}where $1$ is the label of true sample $x_{r}$. $0$ is the label of fake sample $x_{f}=G(z)$. $\phi$ and $\theta$ are the trainable parameters of $G$ and $D$ respectively. Note that when $G$ is trained, $D$ is untrainable. Interested readers can refer to \cite{DBLP:journals/spm/CreswellWDASB18},\cite{DBLP:journals/ieeejas/WangGDLZW17} for surveys of GAN.

\subsubsection{\textbf{GAN in Traffic Domain}}
When GAN is applied in traffic prediction tasks \cite{DBLP:journals/tits/LinDLW19},\cite{liang2018deep}, Generator $G$ is usually employed to generate future traffic observations based on the historical observations. Then the generated data and the future real data are fed into Discriminator $D$ to train it. After training, Generator $G$ can learn the distribution of the real traffic flow data through a large number of historical data and can be used to predict the future traffic states \cite{DBLP:conf/ijcnn/ZhangWCC19}. GAN can be also utilized to solve the sparsity problem of traffic data for its efficacy in handling data generation \cite{DBLP:journals/tits/YuG19}.

In addition, the generator or discriminator of GAN can be any model, such as RNNs, Seq2Seq, depending on the specific traffic tasks.

\section{Challenges Perspective}
\label{sec:Challenges}
Traffic tasks are very challenging due to the complicated spatial dependency, temporal dependency in traffic data. In addition, external factors such as holiday or event can also affect the traffic conditions. In this section, we introduce four common challenges in traffic domain. We carefully examine each challenge and its corresponding solutions, making necessary comparison.

\subsection{Spatial Dependency}
%%%%%%%%%%%%%%%%%%%%% Figures %%%%%%%%%%%%%%%%%%
\begin{figure}[htb]
\centering
\includegraphics[width=0.3\textwidth, height=0.15\textheight]{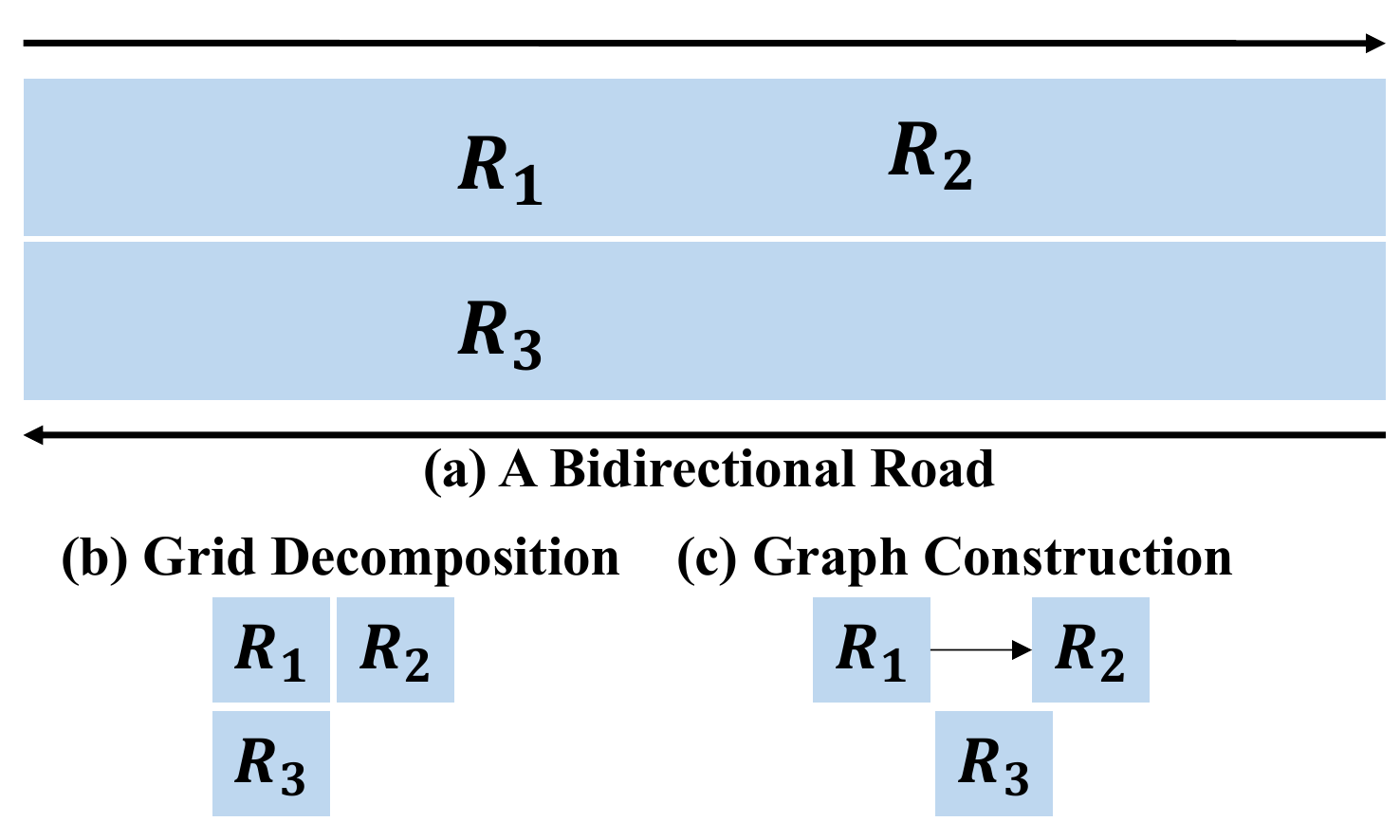}
\caption{The formulation of a bidirectional road: The traffic condition of road $R_1$ is only influenced by the same side road $R_2$ and has weak correlation with the opposite side road $R_3$. But if this region is modeled as grids, $R_3$ has similar impact on $R_1$ as $R_2$, which is against the truth. If it is model as a graph,  $R_1$ is connected with $R_2$ and disconnected with $R_3$, which can reflect the true relationship.}
\label{fig:Spatial Dependency}
\end{figure}
As mentioned in previous section, some literatures\cite{zhang2017deep},\cite{DBLP:journals/ijcisys/DuLGH20},\cite{yao2018deep} extract spatial features through decomposing the whole traffic network into grids and then employing CNNs to process the grid-based data. However, the grid-based assumption actually violates the nature topology of traffic network. Many traffic networks are physically organized as a  graph and the graph topology information is obviously valuable for traffic prediction (as shown in Figure \ref{fig:Spatial Dependency}).  According to our survey, graph neural networks can model spatial dependencies in graph-based traffic networks much better  than grid-based approaches. In addition, the complicated spatial dependencies in traffic network can be categorized into three spatial attributes, i.e. spatial locality, multiple relationships and global connectivity. Different kinds of GNNs combining with other deep learning techniques are utilized to solve different kinds of spatial attributes.

\subsubsection{\textbf{Spatial Locality}}
Spatial locality refers that adjacent regions are usually highly relevant to each other. For example, the passenger flow of a station in a subway is obviously affected by its connected stations. $\mathbf{K}$-localized spectral graph convolution network (SGCN) is widely adopted to aggregate the information of $0$ to $\mathbf{K}-1$ hop neighbors to the central region. In addition, some works make different assumptions about the spatial locality and utilize some novel tricks.

The adjacency matrix representing the traffic topology is usually pre-defined while some works \cite{DBLP:conf/aaai/GuoLFSW19},\cite{chen2019multi} argued that neighboring locations are dynamically correlated with each other. They incorporated the attention mechanism into SGCN to adaptively capture the dynamic correlations among surrounding regions. 

SGCN requires all the regions to have the same local statistics and its convolution kernel is location-independent. However, Zhang et al. \cite{DBLP:conf/icpr/ZhangJCXP18} clarified that the local statistics of traffic data changed from region to region and they designed location-dependent kernels for different regions automatically.

\subsubsection{\textbf{Multiple Relationships}}
While locality attribute focuses on spatial proximity, the target region can be correlated with distant regions through various non-Euclidean relationships such as functional similarity, transportation connectivity (as shown in Figure \ref{fig:Multi-relationships}), semantic neighbors. Functional similarity refers that distant region is similar to the target region in terms of functionality, which can be characterized by the surrounding POIs \cite{geng2019spatiotemporal},\cite{DBLP:conf/mdm/GeLLZ19}. Transportation connectivity suggests that those geographically distant but conveniently reachable can be correlated \cite{geng2019spatiotemporal}. The reachable way can be motorway, highway, subway. Semantic neighbors are adopted to model the correlation between origins and destinations \cite{DBLP:conf/kdd/WangYCW0019}. The correlation is measured by the passenger flow between them. To explicitly extract these correlation information,  different types of correlations using multiple graphs are encoded \cite{geng2019spatiotemporal} and  multi-graph convolution is leveraged. 

\subsubsection{\textbf{Global Connectivity}}
Both spatial proximity and multi-relationship focus on parts of the network while ignore the whole structure. Global connectivity refers that traffic conditions of different regions have influenced each other at a whole network scale. There are several strategies to exploit the global structure information of traffic network.

A popular way to capture global connectivity is to model the changing traffic conditions in the traffic network as a diffusion process that happens at a network scale, which is presented by a power series of transition matrices. Then, diffusion graph convolution network (DGCN) is adopted to extract the spatial dependency globally \cite{DBLP:conf/aaai/ChenLTZWWZ19}, \cite{DBLP:conf/iclr/LiYS018}, \cite{DBLP:conf/ijcnn/ZhangWCC19}, \cite{DBLP:conf/ijcai/WuPLJZ19}, \cite{DBLP:conf/trustcom/HuangWYC19},\cite{DBLP:journals/corr/abs-1906-00560}.

A novel spatial graph pooling layer with path growing algorithm is designed to produce a coarser graph \cite{DBLP:journals/corr/abs-1903-05631}. This pooling layer is stacked before SGC layer to get multi-granularity graph convolutions, which can extract spatial features at various scopes.

A SGC layer with a self-adaptive adjacency matrix  is proposed \cite{DBLP:conf/ijcai/WuPLJZ19} to capture the hidden global spatial dependency in the data. This self-adaptive adjacency matrix is learned from the data through an end-to-end supervised training.

\subsection{Temporal Dependency}
Temporal dependency refers that prediction of traffic conditions at a certain time is usually correlated with various historical observations \cite{DBLP:conf/ijcai/YuYZ18}.  

As stated in Section \ref{sec:Techniques}, many works extract the temporal dependency by RNNs-based approaches. However, RNNs-based approaches suffer from time-consuming iterations and confront gradient vanishing/explosion problem for capturing long sequence. Compared with RNNs-based approaches, TCN-based approaches have the superiority of simple structures, parallel computing and stable gradients. Therefore, some works \cite{DBLP:conf/ijcai/YuYZ18},\cite{DBLP:conf/mdm/GeLLZ19} adopt TCN-based approaches to capture the temporal pattern in traffic data.
In addition, TCN is able to handle different temporal levels by stacking multiple layers. For instance, Fang et al. \cite{DBLP:conf/ijcai/FangZMXP19} and Wu et al. \cite{DBLP:conf/ijcai/WuPLJZ19} stacked multiple TCN layers with the bottom layers extracting short-term neighboring dependencies and the higher layers learning long-term temporal patterns.

\subsubsection{\textbf{Multi-timescale}}
Some works extract the temporal dependency at a multi-timescale perspective \cite{DBLP:conf/aaai/GuoLFSW19},\cite{DBLP:journals/corr/abs-1903-07789}. Temporal dependency is decomposed into recent, daily and weekly dependencies \cite{DBLP:conf/aaai/GuoLFSW19}. The recent dependency refers that the future traffic conditions are influenced by the traffic conditions recently. For instance, the traffic congestion at 9 am inevitably influences traffic flow at the following hours. Daily dependency describes that the repeated daily pattern in traffic data due to the regular daily routine of people, such as morning peak and evening peak. Weekly dependency considers the influence caused by the same week attributes. For instance, all Mondays share similar traffic pattern in a short-term. Guo et al. \cite{DBLP:conf/aaai/GuoLFSW19} set three parallel components with the same structure to model these three temporal attributes respectively.

\subsubsection{\textbf{Different Weights}}
Some works argue that the correlations between historical and future observations are varying at different previous time slices. Guo et al. \cite{DBLP:conf/aaai/GuoLFSW19} adopted a temporal attention mechanism to adaptively attach different importance to historical data.

\subsection{Spatiotemporal Dependency}
Many works capture the spatial and temporal dependency separately in a sequential manner \cite{zhang2019multistep}, \cite{DBLP:conf/ijcnn/ZhangWCC19}, \cite{cuitraffic19}, \cite{DBLP:journals/corr/abs-2002-00786},\cite{DBLP:journals/corr/abs-1912-01242},\cite{DBLP:journals/corr/abs-1909-07105},\cite{DBLP:journals/corr/abs-2003-11973} while the spatial and temporal dependencies are closely intertwined in traffic data. Guo et al. \cite{DBLP:conf/aaai/GuoLFSW19} argued that the historical observations in different locations at different times have varying impacts on central region in the future. Take an obvious example, a traffic accident in a critical road results in serious disruptions over related roads but at different time, due to the gradual formation and dispersion of traffic congestion.

A limitation of separately modeling is that the potential interactions between spatial features and temporal features are ignored, which may hurt the prediction performance. To overcome such limitation, a popular way is to incorporate the graph convolution operations (e.g. SGC, DGC) to RNNs (as stated in Section \ref{sec:Challenges}) to capture spatial-temporal correlations jointly \cite{DBLP:conf/kdd/LiHCSWZP19}, \cite{DBLP:conf/aaai/ChenLTZWWZ19}, \cite{DBLP:conf/iclr/LiYS018}, \cite{guooptimized20}, \cite{DBLP:conf/trustcom/HuangWYC19},\cite{DBLP:journals/corr/abs-1906-00560},\cite{chen2019multi}, \cite{DBLP:journals/corr/abs-1903-05631}, \cite{DBLP:journals/corr/abs-2001-04889}.

\subsection{External Factors }
Factors such as holidays, time attributes (e.g. hour, day, week, month, season, year) \cite{DBLP:conf/mdm/GeLLZ19},\cite{DBLP:journals/corr/abs-1903-07789}, weather (e.g. rainfall, temperature, air quality)\cite{DBLP:journals/corr/abs-1903-07789}, special events, POIs\cite{geng2019spatiotemporal} and traffic incidents (e.g. incident time, incident type) \cite{DBLP:journals/corr/abs-1912-01242} can influence the traffic prediction in some extent, which we refer as external factors or context factors. In addition, Zhang et al. \cite{zhang2019multistep} considered historical statistical speed information (e.g. average or standard deviation of traffic speed) as external factor.

Some factors such as day attributes, holidays and weather conditions are encoded as discrete values and they are usually transformed into binary vectors by one-hot encoding. Other factors including temperature, wind speed are encoded as  continual values  and they are usually normalized by Min-Max normalization or Z-score normalization.

There are two approaches to handle  external factors in the literatures we survey. The first approach is to concatenate the external factors with other features and feed them into model \cite{DBLP:conf/aaai/ChenLTZWWZ19}, \cite{DBLP:conf/mdm/GeLLZ19}. The second approach is to design an external component in charge of processing external factors alone. The external component usually contains two fully connected layers, of which the first extracting important features and the second mapping low dimension features to high dimension features \cite{DBLP:conf/mdm/GeLLZ19}, \cite{DBLP:journals/corr/abs-1912-01242},\cite{DBLP:journals/corr/abs-1903-07789},\cite{DBLP:conf/gis/ChaiWY18}. Bai et al. \cite{DBLP:conf/cikm/BaiYK0LY19} employed multi-LSTM layers to extract representation of external factors. The output of external component is fused with other components to generate the final result.

\section{Public Datasets and Open Source Codes}
\label{sec:Open}
%%%%%%%%%%%%%%%%%%    Table  %%%%%%%%%%%%%%%%%
\begin{table*}[htb]
\caption{Some open traffic datasets}
\centering
\scriptsize
\begin{tabular}{c|c|c}
\toprule
\textbf{Datasets} &\textbf{Links}  &\textbf{References}\\
\midrule
NYC taxi&https://www1.nyc.gov/site/tlc/about/tlc-trip-record-data.page&\cite{DBLP:conf/aaai/Diao0ZLXH19}, \cite{DBLP:journals/corr/abs-1903-07789},\cite{DBLP:journals/corr/abs-1912-01242},\cite{DBLP:journals/corr/abs-1910-09103}\\\hline
NYC bike&https://www.citibikenyc.com/system-data&\cite{DBLP:journals/corr/abs-1903-07789}, \cite{DBLP:conf/gis/ChaiWY18}, \cite{DBLP:conf/ijcai/BaiYK0S19}, \cite{DBLP:conf/cikm/BaiYK0LY19}\\\hline
San Francisco taxi&https://crawdad.org/~crawdad/epfl/mobility/20090224/&\cite{DBLP:journals/corr/abs-1912-01242}\\\hline
Chicago bike&https://www.divvybikes.com/system-data&\cite{DBLP:conf/gis/ChaiWY18}\\\hline
BikeDC (Bike Washington)&https://www.capitalbikeshare.com/system-data&\cite{DBLP:journals/corr/abs-1903-07789}\\\hline
California -PEMS&http://pems.dot.ca.gov/&\cite{DBLP:conf/ijcai/YuYZ18},\cite{DBLP:conf/mdm/GeLLZ19},\cite{DBLP:conf/aaai/GuoLFSW19},\cite{DBLP:conf/aaai/Diao0ZLXH19},\cite{DBLP:conf/aaai/ChenLTZWWZ19},\cite{DBLP:journals/corr/abs-1903-00919},\cite{DBLP:journals/corr/abs-1911-08415},\cite{DBLP:conf/ijcai/WuPLJZ19},\cite{guooptimized20},\cite{DBLP:conf/kdd/LiHCSWZP19},\cite{DBLP:conf/trustcom/HuangWYC19}\\
\bottomrule
\end{tabular}
\label{tab:datasets}
\end{table*}

\subsection{Public Datasets}
We summarize some public datasets (as shown in Table \ref{tab:datasets}) in the literatures we survey to help successors participate in this domain and produce more valuable works. 

\subsection{Open Source Codes}
Open-source implementations are helpful for researchers to compare their approaches. We provide the hyperlinks of public source codes of the literatures reviewed in this paper (as shown in Table  \ref{tab:codes}) to facilitate the baseline experiments in traffic domain.
%%%%%%%%%%%%%%%%%%%%%%% Table %%%%%%%%%%%%%
\begin{table*}[htb]
\caption{Some open source codes}
\centering
\scriptsize
\begin{tabular}{c|c|c|c|c}
\toprule
\textbf{Reference} &\textbf{Model}  &\textbf{Year} &\textbf{Framework} &\textbf{Github}\\ 
\hline
\cite{DBLP:conf/iclr/LiYS018}&DCRNN&2018&Tensorflow&https://github.com/liyaguang/DCRNN\\\hline
\cite{DBLP:conf/uic/LiPLXDMWB18}&GCNN&2018&Keras&https://github.com/RingBDStack/GCNN-In-Traffic\\\hline
\cite{zhaoTGCN19}&T-GCN&2019&Tensorflow&https://github.com/lehaifeng/T-GCN\\\hline
\cite{DBLP:journals/corr/abs-1911-08415}&GMAN&2019&Tensorflow&https://github.com/zhengchuanpan/GMAN\\\hline
\cite{DBLP:conf/ijcai/WuPLJZ19}&Graph-WaveNet&2019&Torch&https://github.com/nnzhan/Graph-WaveNet\\
\bottomrule
\end{tabular}
\label{tab:codes}
\end{table*}

\section{Future Directions}
\label{sec:Future}
We have investigated the latest advances in graph-based traffic literatures and made a summary of these literatures in Table \ref{tab:architectures}. Further, we suggest some directions for researchers to explore, which can be divided into three categories, i.e. application related, technique related, external factor related directions.

\subsubsection{\textbf{Application Related Directions}}
As shown in Table \ref{tab:architectures}, there are many works utilizing graph-based deep learning architectures to tackle traffic state prediction and traffic demand prediction, which have achieved state-of-the-art performance. However, there are only a handful of works analyzing traffic data in a graph perspective in other research directions, such as vehicle behavior classification \cite{DBLP:journals/corr/abs-2002-00786}, optimal dynamic electronic toll collection (DETC) scheme \cite{qiu2019dynamic}, path availability \cite{DBLP:conf/kdd/LiHCSWZP19}, traffic signal control \cite{DBLP:conf/itsc/NishiOHY18}. When it comes to traffic incident detection, vehicle detection, origin-destination travel demand prediction and transfer learning from City to City, works adopting graph-based deep learning techniques are rare up to now. Therefore, the upcoming participators can explore these directions in a graph perspective and learn the successful experiences from existing works.

\subsubsection{\textbf{Technique Related Directions}} On one hand, most existing works have employed spectral graph convolution network (SGCN) and diffusion graph convolution network (DGCN), two popular kinds of GNNs, to analyze traffic tasks. There are only a handful of works utilizing Graph attention networks (GATs) in traffic domain \cite{DBLP:conf/iclr/VelickovicCCRLB18}, \cite{DBLP:journals/corr/abs-1911-08415}, \cite{kanglearning19}, \cite{DBLP:journals/access/ZhangYL19a},\cite{DBLP:conf/uai/ZhangSXMKY18}. Other kinds of GNNs, such as graph auto-encoders (GAEs) \cite{DBLP:conf/nips/HasanzadehHNDZQ19},\cite{simonovsky2018graphvae}, recurrent graph neural networks (RecGNNs) \cite{dai2018learning} have achieved state-of-the-art performance in other domains, but they are seldom explored in traffic domain up to  now. Therefore, it is worth to extend these branches of GNNs to traffic domain. On the other hand, recent works have combined GNNs with other deep learning techniques such as RNNs, TCN, Seq2Seq, GAN to solve the challenges in traffic tasks. However, few traffic works consider transfer learning, continue learning and reinforcement learning together with GNNs, which might be a promising direction for researchers. In addition, most of the graph-based traffic works are regression tasks, while classification tasks are few \cite{DBLP:conf/kdd/LiHCSWZP19},\cite{DBLP:journals/corr/abs-2002-00786}. Researchers can explore the classification traffic tasks in a graph perspective.

\subsubsection{\textbf{External Factors Related Directions}}
Finally, many existing traffic models do not take external factors into consideration, for that external factors are hard to collect and have various formats. The data sparsity of external factors is still a challenge confronted by the research community. In addition, the techniques to process external factors are rather naive, e.g. a simple fully connected layer. There should be more approaches to process external factors.

\section{Conclusion}
\label{sec:Conclusion}
In this survey, we conduct a comprehensive review of various graph-based deep learning architectures in recent traffic works. More specifically, we summarize a general graph-based formulation of traffic problem and graph construction from various traffic datasets. Further, we decompose all the investigated architectures and analyze the common modules they share, including graph neural networks (GNNs), recurrent neural networks (RNNs),  temporal convolution network (TCN), Sequence to Sequence (Seq2Seq) model, generative adversarial network (GAN). We provide a thorough description of their variants in traffic tasks, hoping to provide upcoming researchers insights into how to design novel techniques for their own traffic tasks. We also summarize the common challenges in many traffic scenarios, such as spatial dependency, temporal dependency, external factors. More than that, we present multiple deep learning based solutions for each challenge. In addition, we provide some hyperlinks of public datasets and codes in related works to facilitate the upcoming researches. Finally, we suggest some future directions for participators interested in this domain.

\section*{Acknowledgment}
The authors would like to thank anonymous reviewers for their valuable comments.

This work is supported by the National Key R\&D Program of China (No.2019YFB2102100), National Natural Science Foundation of China (No.61802387), China’s Post-doctoral Science Fund (No.2019M663183), National Natural Science Foundation of Shenzhen (No.JCYJ20190812153212464), Shenzhen Engineering Research Center for Beidou Positioning Service Improvement Technology  (No.XMHT20190101035), Science and Technology Development Fund of Macao S.A.R (FDCT) under number 0015/2019/AKP, Shenzhen Discipline Construction Project for Urban Computing and Data Intelligence.
\bibliographystyle{IEEEtran}
\bibliography{IEEEabrv,reference}

\begin{IEEEbiography}[{\includegraphics[width=1in,height=1.25in,clip,keepaspectratio]{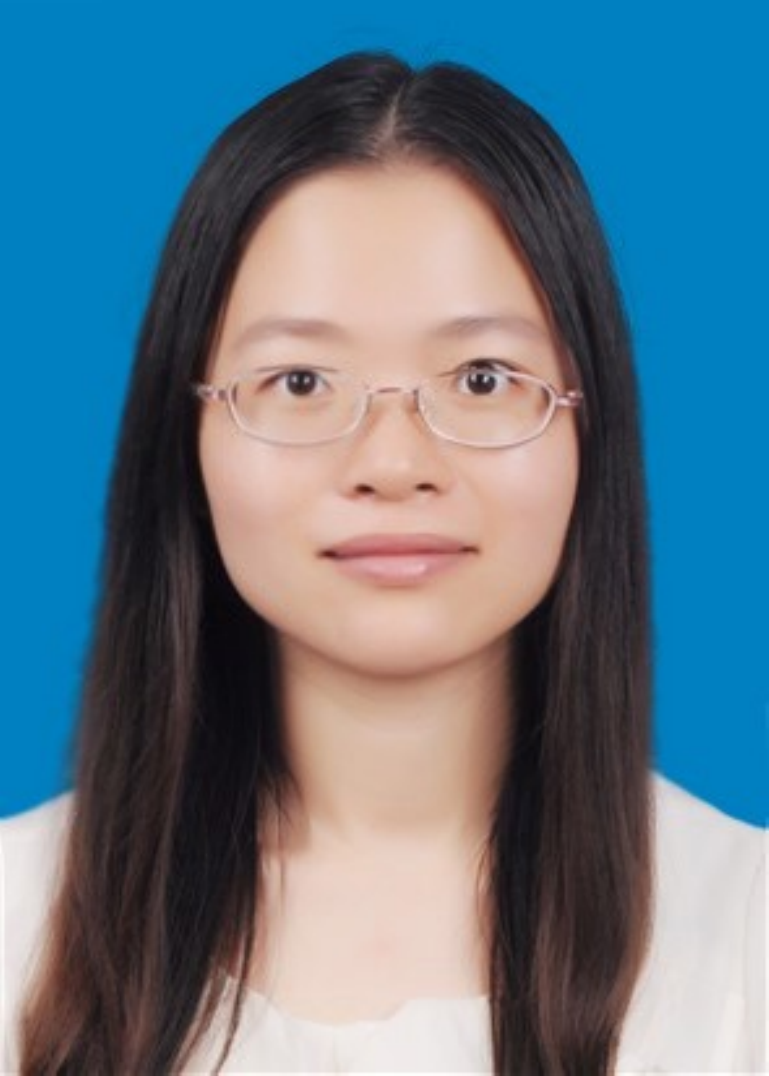}}]{Jiexia Ye}
received the Bachelor's degree in Economics from Sun Yat-sen University in 2012. She is currently working toward M.S. degree in Shenzhen Institutes of Advanced Technology, Chinese Academy of Sciences. Her research interests include graph neural networks / graph embedding in traffic and finance domain.
\end{IEEEbiography}
\begin{IEEEbiography}[{\includegraphics[width=1in,height=1.25in,clip,keepaspectratio]{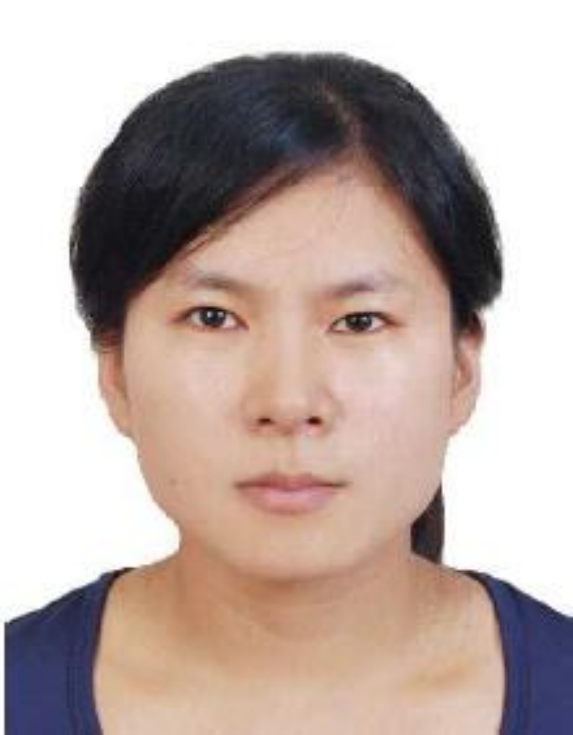}}]{Juanjuan Zhao} received her Ph.D degree from Shenzhen College of Advanced Technology, University of Chinese Academy of Sciences in 2017, and received the M.S. degree from the Department of Computer Science, Wuhan University of Technology in 2009. She is an Assistant Professor at Shenzhen Institutes of Advanced Technology, Chinese Academy of Sciences. Her research topics include data-driven urban systems, mobile data collection, cross-domain data fusion, heterogeneous model integration.
\end{IEEEbiography}
\begin{IEEEbiography}[{\includegraphics[width=1in,height=1.25in,clip,keepaspectratio]{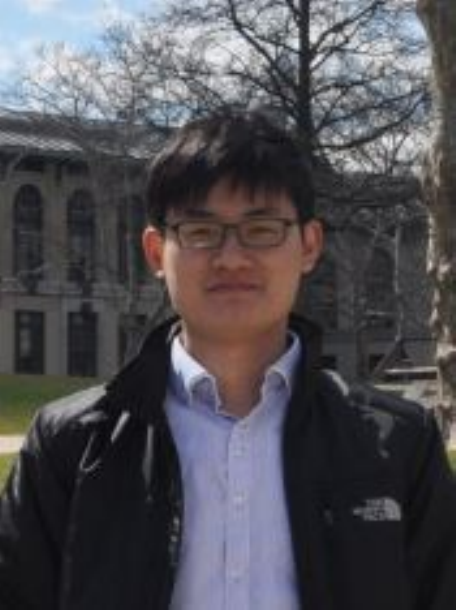}}]{Kejiang Ye} received his BSc and Ph.D degree in Computer Science from Zhejiang University in 2008 and 2013 respectively. He was also a joint Ph.D student at The University of Sydney from 2012 to 2013. After graduation, he worked as Post-Doc Researcher at Carnegie Mellon University from 2014 to 2015 and Wayne State University from 2015 to 2016. He is currently an Associate Professor at Shenzhen Institutes of Advanced Technology, Chinese Academy of Science. His research interests include cloud computing, big data and network systems.
\end{IEEEbiography}
\begin{IEEEbiography}[{\includegraphics[width=1in,height=1.25in,clip,keepaspectratio]{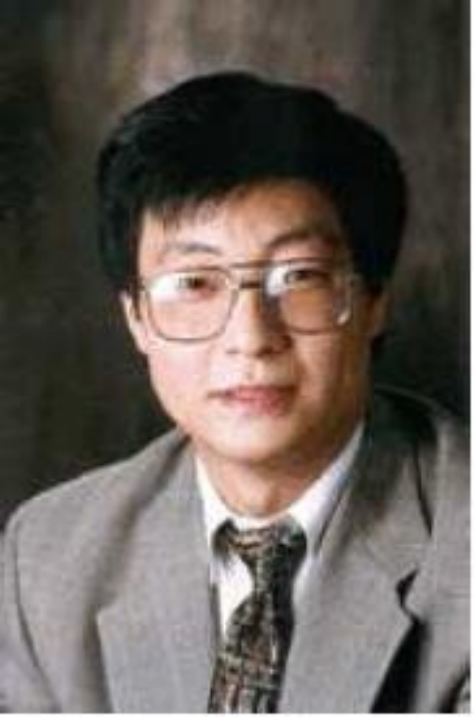}}]{Chengzhong Xu} received his Ph.D degree from the University of Hong Kong, China in 1993. He is the Dean of the Faculty of State Key Lab of IOTSC, Department of Computer Science, University of Macau, Macao SAR, China and a Chair Professor of Computer Science of UM. He was a Chief Scientist of Shenzhen Institutes of Advanced Technology (SIAT) of Chinese Academy of Sciences and the Director of Institute of Advanced Computing and Digital Engineering of SIAT.  He was also in the faculty of Wayne State University, USA for 18 years. Dr. Xu's research interest is mainly in the areas of parallel and distributed systems, cloud and edge computing, and data-driven intelligence. He has published over 300 peer-reviewed papers on these topics with over 10K citations. Dr. Xu served in the editorial boards of leading journals, including IEEE Transactions on Computers, IEEE Transactions on Cloud Computing, IEEE Transactions on Parallel and Distributed Systems and Journal of Parallel and Distributed Computing. He is the Associate Editor-in-Chief of ZTE Communication. He is IEEE Fellow and the Chair of IEEE Technical Committee of Distributed Processing.
\end{IEEEbiography}

\end{document}